\documentclass[journal,nofonttune]{IEEEtran}

\usepackage{fontspec}

\usepackage{breqn}
\usepackage{lineno}
\usepackage{setspace}
\usepackage{here}
\usepackage{bbm}
\usepackage{fancyhdr}
\usepackage[margin=15mm]{geometry}
\usepackage{amsmath,amssymb}
\usepackage{graphicx}
\usepackage{xcolor}
\usepackage{booktabs}
\usepackage{url}
\usepackage{cancel}
\usepackage{tikz}
\usepackage[most]{tcolorbox}
\usepackage{todonotes}
\tcbuselibrary{skins, breakable}
\usepackage[caption=false,font=footnotesize]{subfig}
\usepackage{amsthm}
\usepackage{enumitem}
\usepackage{aliascnt}
\theoremstyle{plain}

\usepackage{unicode-math}
\newcommand{\Fkx}{\{0,1\}^k\setminus\{{\bf 0}\}}
\allowdisplaybreaks

\newtheorem{theorem}{Theorem}

\newaliascnt{corollary}{theorem}
\newtheorem{corollary}[corollary]{Corollary}
\aliascntresetthe{corollary}

\newaliascnt{lemma}{theorem}
\newtheorem{lemma}[lemma]{Lemma}
\aliascntresetthe{lemma}

\newaliascnt{conjecture}{theorem}
\newtheorem{conjecture}[conjecture]{Conjecture}
\aliascntresetthe{conjecture}

\newaliascnt{proposition}{theorem}
\newtheorem{proposition}[proposition]{Proposition}
\aliascntresetthe{proposition}

\theoremstyle{definition}

\newaliascnt{definition}{theorem}
\newtheorem{definition}[definition]{Definition}
\aliascntresetthe{definition}

\theoremstyle{definition}

\newaliascnt{remark}{theorem}
\newtheorem{remark}[remark]{Remark}
\aliascntresetthe{remark}

\usepackage[createShortEnv]{proof-at-the-end}
\newEndThm{remarkE}{remark}

\tcbset{
  apxthmstyle/.style={
    enhanced,
    breakable,
    colback=black!2,
    colframe=black!45,
    boxrule=0.4pt,
    arc=1pt,
    left=6pt,
    right=6pt,
    top=4pt,
    bottom=4pt,
    before skip=8pt,
    after skip=8pt,
  },
  apxproofstyle/.style={
    enhanced,
    breakable,
    colback=white,
    colframe=black!30,
    boxrule=0.4pt,
    arc=1pt,
    left=6pt,
    right=6pt,
    top=4pt,
    bottom=4pt,
    before skip=6pt,
    after skip=8pt,
  }
}

\let\oldtheorem\theorem
\let\endoldtheorem\endtheorem
\RenewDocumentEnvironment{theorem}{o}
  {\begin{tcolorbox}[apxthmstyle]%
   \IfNoValueTF{#1}{\oldtheorem}{\oldtheorem[#1]}}
  {\endoldtheorem\end{tcolorbox}}

\let\oldlemma\lemma
\let\endoldlemma\endlemma
\RenewDocumentEnvironment{lemma}{o}
  {\begin{tcolorbox}[apxthmstyle]%
   \IfNoValueTF{#1}{\oldlemma}{\oldlemma[#1]}}
  {\endoldlemma\end{tcolorbox}}

\let\oldcorollary\corollary
\let\endoldcorollary\endcorollary
\RenewDocumentEnvironment{corollary}{o}
  {\begin{tcolorbox}[apxthmstyle]%
   \IfNoValueTF{#1}{\oldcorollary}{\oldcorollary[#1]}}
  {\endoldcorollary\end{tcolorbox}}

\let\oldproposition\proposition
\let\endoldproposition\endproposition
\RenewDocumentEnvironment{proposition}{o}
  {\begin{tcolorbox}[apxthmstyle]%
   \IfNoValueTF{#1}{\oldproposition}{\oldproposition[#1]}}
  {\endoldproposition\end{tcolorbox}}

\let\olddefinition\definition
\let\endolddefinition\enddefinition
\RenewDocumentEnvironment{definition}{o}
  {\begin{tcolorbox}[apxthmstyle]%
   \IfNoValueTF{#1}{\olddefinition}{\olddefinition[#1]}}
  {\endolddefinition\end{tcolorbox}}

\let\oldremark\remark
\let\endoldremark\endremark
\RenewDocumentEnvironment{remark}{o}
  {\begin{tcolorbox}[apxthmstyle]%
   \IfNoValueTF{#1}{\oldremark}{\oldremark[#1]}}
  {\endoldremark\end{tcolorbox}}

\let\oldconjecture\conjecture
\let\endoldconjecture\endconjecture
\RenewDocumentEnvironment{conjecture}{o}
  {\begin{tcolorbox}[apxthmstyle]%
   \IfNoValueTF{#1}{\oldconjecture}{\oldconjecture[#1]}}
  {\endoldconjecture\end{tcolorbox}}
  
\usepackage{tikz}
\usetikzlibrary{matrix,positioning,backgrounds}
\usepackage{thmtools}
\usepackage[numbers,sort&compress]{natbib}
\usepackage{enumitem}
\usepackage{braket}

\makeatletter
\newif\ifaxp@alreadyappendix
\let\axp@origappendix\appendix
\renewcommand{\appendix}{%
  \global\axp@alreadyappendixtrue
  \axp@origappendix
}
\makeatother

\newcommand{\V}{\{0,1\}^L}

\begin{document}

\title{Random Access Codes: Explicit Constructions, Optimality, and Classical--Quantum Gaps}

\author{
Ruho Kondo,
Yuki Sato,
Hiroshi Yano,
Yota Maeda,
Kosuke Ito,
and Naoki Yamamoto
\thanks{Ruho Kondo is with Toyota Central R\&D Labs., Inc., 41-1 Yokomichi, Nagakute, Aichi 480-1192, Japan, and also with the Quantum Computing Center, Keio University, 3-14-1 Hiyoshi, Kohoku-ku, Yokohama, Kanagawa 223-8522, Japan (e-mail: r-kondo@mosk.tytlabs.co.jp).}
\thanks{Yuki Sato and Hiroshi Yano are with Toyota Central R\&D Labs., Inc., 1-4-14 Koraku, Bunkyo-ku, Tokyo 112-0004, Japan, and also with the Quantum Computing Center, Keio University, 3-14-1 Hiyoshi, Kohoku-ku, Yokohama, Kanagawa 223-8522, Japan.}
\thanks{Yota Maeda is with Toyota Central R\&D Labs., Inc., 1-4-14 Koraku, Bunkyo-ku, Tokyo 112-0004, Japan.}
\thanks{Kosuke Ito is with the Advanced Material Engineering Division, Toyota Motor Corporation, 1200 Mishuku, Susono, Shizuoka 410-1193, Japan, and also with the Quantum Computing Center, Keio University, 3-14-1 Hiyoshi, Kohoku-ku, Yokohama, Kanagawa 223-8522, Japan.}
\thanks{Naoki Yamamoto is with the Quantum Computing Center, Keio University, 3-14-1 Hiyoshi, Kohoku-ku, Yokohama, Kanagawa 223-8522, Japan, and also with the Department of Applied Physics and Physico-Informatics, Keio University, Hiyoshi 3-14-1, Kohoku-ku, Yokohama 223-8522, Japan.}
}

\maketitle

\begin{abstract}
A random access code (RAC) encodes an $L$-bit string into a $k$-bit message, $L>k$, so that any requested bit can be recovered with high probability; a quantum RAC (QRAC) uses $k$ qubits instead. We give a geometric characterization of optimal classical $(L,k)$-RACs under average and worst-case decoding criteria. The average criterion is reduced to choosing $2^k$ representatives in $\{0,1\}^L$, while the worst-case criterion is reduced to a minimax problem over $2^k$ points in $[0,1]^L$ with a distance-like objective. This framework proves optimality for several parameter families, with many optimal constructions arising from standard infinite families of binary linear codes. It also yields two explicit classical--quantum separations. First, for every $L>1$, we construct a $(L,1)$-QRAC whose average decoding success probability strictly exceeds the optimal classical value. Second, for the family $(2^k-1,k)$, we prove worst-case optimality of a classical RAC and construct a QRAC with strictly larger worst-case success probability. For the family $(L,L-1)$, the framework identifies a classical RAC that is average-case optimal and, under a stated conjecture, also worst-case optimal. The same viewpoint further recovers explicit $(L,L-1)$-QRACs attaining a previously conjectured upper-bound value.
\end{abstract}



\section{Introduction}
\label{sec:introduction}

\begin{table*}[tbh]
\centering
\caption{
Known exact values and bounds on the maximum decoding success probabilities of $(L,k)$-(Q)RACs under each criterion.
Here, C and Q denote the classical and quantum settings, respectively, and $L,k$ are positive integers satisfying $k<L$.
The function $\mathcal{H}_2(p)=-p\log_2 p-(1-p)\log_2(1-p)$ denotes the binary entropy function.
}
\label{tab:summary-known}
\renewcommand{\arraystretch}{1.3}
\begin{tabular}{ccllc}
\toprule
Type & Criterion & $(L,k)$ & Success Prob. & Reference \\
\midrule
C & Average & $(L,1)$ & $\frac{1}{2}+\frac{1}{2^L}\binom{L-1}{\left\lfloor L/2\right\rfloor}$ & \cite{ambainis2024quantum} \\
C & Worst-case & $(L,k)$ & $\le\mathcal{H}_2^{-1}\big(1-\frac{k}{L}\big)$ & \cite{ambainis1999dense} \\
C & Both & $(L,k)$ & $\ge\frac{1}{2}+\frac{1}{2}\sqrt{\frac{1}{(2^k-1)L}}$ & \cite{liabotro2017improved} \\
\midrule
Q & Both & $(2,1)$ & $\frac{1}{2}+\frac{1}{2}\sqrt{\frac{1}{2}}$ & \cite{ambainis1999dense}\\
Q & Both & $(3,1)$ & $\frac{1}{2}+\frac{1}{2}\sqrt{\frac{1}{3}}$ & \cite{ambainis1999dense}\\
Q & Both & $(3,2)$ & $\frac{1}{2}+\frac{1}{2}\sqrt{\frac{2}{3}}$ & \cite{imamichi2018constructions}\\
Q & Worst-case & $(L,k)$ & $\le\mathcal{H}_2^{-1}\big(1-\frac{k}{L}\big)$ & \cite{nayak1999optimal} \\
Q & Average & $(L,k)$ & $\le\frac{1}{2}+\frac{1}{2}\sqrt{\frac{2^{k-1}}{L}}$ & \cite{manvcinska2022geometry}\\
Q & Both & $(L,k)$ & $\ge\frac{1}{2}+\frac{1}{2}\sqrt{\frac{1}{(2^k-1)L}}$ & \cite{liabotro2017improved} \\
Q & Both & $(L,L-1)$ & $\ge\frac{1}{2}+\frac{1}{2}\sqrt{\frac{L-1}{L}}$ & \cite{suzuki2026analytical}\\
\bottomrule
\end{tabular}
\end{table*}

\begin{table*}[tbh]
\centering
\caption{
Summary of analytically obtained decoding success probabilities for the families of classical RACs and QRACs considered in this paper, together with corresponding classical codes when available.
Here, C and Q denote the classical and quantum settings, respectively, and $k,L,m$ are positive integers with $L,m\ge2$.
The ``Optimality'' column specifies whether the listed value is proved optimal (Proved), optimal conditional on Conjecture~\ref{conjecture:worst} (Conditional), or only known to be achievable by an explicit construction (--).
}
\label{tab:summary}
\renewcommand{\arraystretch}{1.3}
\setlength{\tabcolsep}{4pt}
\begin{tabular}{cclllll}
\toprule
Type & Criterion & $(L,k)$ & Success prob. & Optimality & Corresponding code & Ref. \\
\midrule
C & Average & $(2^m-1,2^m-m-1)$ 
& $1-\frac{1}{L+1}$ 
& Proved 
& Hamming 
& Cor.~\ref{corollary:hamming-code} \\
C & Average & $(L,1)$ 
& $\frac{1}{2}+\frac{1}{2^L}\binom{L-1}{\left\lfloor L/2\right\rfloor}$ 
& Proved 
& Repetition 
& Cor.~\ref{corollary:average-l1}, \ref{corollary:repetition-code}\\
C & Average & $(L,L-1)$ 
& $1-\frac{1}{2L}$ 
& Proved 
& Single parity-check
& Cor.~\ref{corollary:average}, \ref{corollary:single-parity-check-avg}\\
C & Worst-case & $(2^k-1,k)$ 
& $\frac{1}{2}+\frac{1}{2L}$ 
& Proved 
& Simplex 
& Thms.~\ref{theorem:worst-2k1}, \ref{theorem:worst-2k1-b}\\
C & Worst-case & $(L,L-1)$ 
& $1-\frac{1}{L}$ 
& Conditional 
& Single parity-check
& Cor.~\ref{theorem:worst2}, \ref{corollary:single-parity-check}\\
\midrule
Q & Average & $(L,1)$
& See Thm.~\ref{theorem:L1qrac}
& --
& --
& Thm.~\ref{theorem:L1qrac}\\
Q & Both & $(2^k-1,k)$ 
& $\frac{1}{2}+\frac{1}{2\sqrt{L}}$ 
& -- 
& --
& Thm.~\ref{theorem:worst-2k1-Q}\\
Q & Both & $(L,L-1)$ 
& $\frac{1}{2}+\frac{1}{2}\sqrt{\frac{L-1}{L}}$ 
& -- 
& --
& Thm.~\ref{theorem:optimal-qrac}\\
\bottomrule
\end{tabular}
\end{table*}

\begin{figure*}[t]
  \centering
    \includegraphics[scale=1]{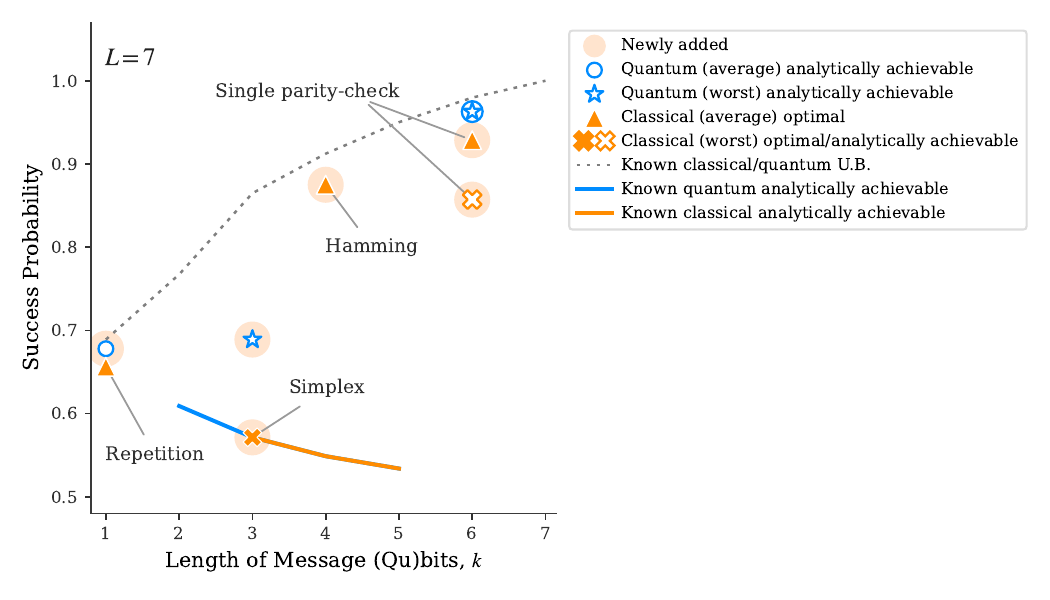}
  \caption{
    Known and newly obtained optimal/achievable decoding success probabilities for $(L,k)$-(Q)RACs.
    Known upper bounds by \cite{manvcinska2022geometry} ($k=1,2$) and \cite{ambainis1999dense,nayak1999optimal} ($k\ge3$) are also shown.
    }
      \label{fig:Lk-QRAC}
\end{figure*}
A random access code (RAC) is a protocol that encodes multiple bits of information into a message shorter than the original information, such that a receiver can probabilistically recover any designated bit position from the message.
When the message is encoded into a quantum state, the protocol is called a quantum random access code (QRAC).
The idea dates back to Wiesner's conjugate coding~\cite{wiesner1983conjugate} and was later rediscovered and formulated explicitly as a random access code by Ambainis {\it et al.}~\cite{ambainis1999dense}.
By convention, a protocol that encodes $L$ bits into a $k$-(quantum)bit message is referred to as an $(L,k)$-(Q)RAC.
In~\cite{ambainis1999dense}, $(2,1)$-QRACs and $(3,1)$-QRACs were proposed, and it is known that they achieve higher decoding success probabilities than the corresponding classical $(2,1)$-RAC and $(3,1)$-RAC.
Leveraging their compressibility, QRACs have also been applied to accelerating combinatorial optimization problems~\cite{fuller2024approximate,teramoto2023role,tamura2024noise,kondo2025recursive,sharma2024quantum,he2025non,matsuyama2025sampling}.
In this way, it is known that QRACs can outperform RACs in terms of decoding success probability for certain specific choices of $L$ and $k$.
However, for general $L$ and $k$, the extent to which a quantum advantage can be achieved remains unclear.

We next summarize several theoretical results known for RACs and QRACs.
First, it has been shown that there exists no $(L,k)$-RAC for $L\ge2^k$ whose worst-case decoding success probability exceeds $\frac{1}{2}$, whereas a $(2^k-1,k)$-RAC does exist~\cite[Corollary~5.3]{iwama2007unbounded}.
Similarly, there exists no $(L,k)$-QRAC for $L\ge4^k$, while a $(4^k-1,k)$-QRAC exists.
Note that there exists an $(L,k)$-(Q)RAC for arbitrary $L,k\in\mathbb{N}$ with whose average decoding success probability exceeds $\frac{1}{2}$.
For the worst-case decoding success probability of an $(L,k)$-RAC, $P^*_{\rm C}(L,k)$, the following upper bound is known~\cite[Theorem~2.1]{ambainis1999dense}:
\begin{equation}
    \label{eq:rac-average-upper}
    P^*_{\rm C}(L,k)\le \mathcal{H}_2^{-1}\left(1-\frac{k}{L}\right).
\end{equation}
Here $\mathcal{H}_2(p)=-p\log_2 p-(1-p)\log_2(1-p)$ denotes the binary entropy function.
Moreover, the worst-case decoding success probability of an $(L,k)$-QRAC, $P^{*}_{\rm Q}(L,k)$, is known to satisfy the same upper bound~\cite[Theorem~2.3]{nayak1999optimal}:
\begin{equation}
    \label{eq:qrac-average-upper}
    P^*_{\rm Q}(L,k)\le \mathcal{H}_2^{-1}\left(1-\frac{k}{L}\right).
\end{equation}
That is, the worst-case decoding success probabilities of RACs and QRACs are known to satisfy the same upper bound.
On the other hand, it has been shown that there exists an $(L,k)$-RAC whose worst-case decoding success probability satisfies
\begin{equation}
    \label{eq:worst-lower}
    P_{\rm C}^*(L,k)\ge \mathcal{H}_2^{-1}\left(1-\frac{k}{L}+\frac{7\log_2 L}{L}\right),
\end{equation}
~\cite[Theorem~2.2]{ambainis1999dense}.
Therefore, in the limit as $L\to\infty$, there exist $(L,k)$-RACs whose worst-case decoding success probability asymptotically attains the upper bound.
This means that the entropy upper bounds \eqref{eq:rac-average-upper} and \eqref{eq:qrac-average-upper} are asymptotically tight, and there are no quantum advantage in the asymptotic regime.
However, the tightness of the entropy bound in the non-asymptotic regime has not been evaluated, and the extent of the gap between the actual supremum of the worst-case decoding success probabilities of RACs and QRACs remains unknown.
Moreover, only a few explicit constructions of RACs and QRACs attaining the upper bound in the non-asymptotic regime are currently known.
More recently, a new upper bound on the average decoding success probability of an $(L,k)$-QRAC, $\overline{P}_{\rm Q}(L,k)$, was established in~\cite[Theorem~10]{manvcinska2022geometry}:
\begin{equation}
    \overline{P}_{\rm Q}(L,k)\le \frac{1}{2}+\frac{1}{2}\sqrt{\frac{2^{k-1}}{L}} .
\end{equation}
Since the optimal worst-case success probability is upper bounded by the optimal average success probability, this also yields an upper bound on the worst-case success probability of $(L,k)$-QRACs.
This new upper bound is tighter than the entropy bounds \eqref{eq:rac-average-upper} and \eqref{eq:qrac-average-upper} for $k=1,2$.
In these cases, the bound can be equivalently expressed as
\begin{equation}
    \label{eq:qrac-conjecture}
    P^*_{\rm Q}(L,k)\le \frac{1}{2}+\frac{1}{2}\sqrt{\frac{k}{L}}
    \quad
    (\mathrm{conjecture}) .
\end{equation}
Motivated by this observation, \cite{manvcinska2022geometry} conjectured that the upper bound in \eqref{eq:qrac-conjecture} holds.
It should be noted, however, that the entropy upper bounds \eqref{eq:rac-average-upper} and \eqref{eq:qrac-average-upper} are asymptotically tight.
Hence, \eqref{eq:qrac-conjecture} can be valid only if it asymptotically matches the entropy upper bounds, which occurs only when either $k=o(L)$ or $L-k=o(L)$.
Upper bounds for general $d$-level $(L,k)$-QRACs were proposed in~\cite[Corollary~1]{farkas2025simple}; however, for the qubit case ($d=2$), the bound in~\cite{manvcinska2022geometry} is tighter.

Although it is known that replacing classical message bits by qubits does not provide an asymptotic advantage in the sense that one cannot substantially reduce the message length $k$~\cite[Theorem~1.2]{ambainis1999dense}, quantum advantage can be observed when focusing on the average success probability in the case $k=1$~\cite{ambainis2024quantum}.
The average decoding success probabilities of the $(2,1)$-RAC and $(3,1)$-RAC have also been discussed in the literature (see, e.g.,~\cite{tavakoli2015quantum}).
In~\cite{ambainis2024quantum}, an average-optimal $(L,1)$-RAC was constructed, and the following upper bound on the average decoding success probability was shown:
\begin{equation}
    \overline{P}_{\rm C}(L,1)\le \frac{1}{2}+\frac{1}{2^L}\binom{L-1}{\lfloor\frac{L}{2}\rfloor} .
\end{equation}
The above results summarize the current theoretical landscape on upper bounds for $(L,k)$-RACs and $(L,k)$-QRACs.

On the other hand, explicit methods to build encoders and decoders for (Q)RACs have also been studied.
In~\cite{imamichi2018constructions}, $(L,2)$-QRACs are constructed via a seesaw semidefinite programming (SDP) algorithm, i.e., an alternating SDP optimization over the quantum states and the positive operator-valued measures (POVMs).
The same work also proposes an optimal $(3,2)$-QRAC in terms of both the average and worst-case decoding success probability, and further discussion of its structure is given in~\cite{teramoto2023quantum}.
A general construction method for $(L<2^k,k)$-RACs and $(L<4^k,k)$-QRACs is discussed in~\cite{liabotro2017improved}, which proposes the following form of encoder:
\begin{equation}
    \label{eq:improved-rho}
    \rho({\bf b})=\frac{1}{2^k}\mathbb{I}^{\otimes k}
    +\frac{1}{2^k\sqrt{(2^k-1)L}}\sum_{i\in[L]}(-1)^{b_i}\xi_i .
\end{equation}
Here $\xi_i$ are Pauli operators; for RACs one uses $\xi_i\in\{\mathbb{I},\texttt{Z}\}^{\otimes k}\setminus\{\mathbb{I}^{\otimes k}\}$, whereas for QRACs one uses $\xi_i\in\{\mathbb{I},\texttt{X},\texttt{Y},\texttt{Z}\}^{\otimes k}\setminus\{\mathbb{I}^{\otimes k}\}$, where $\texttt{X}$, $\texttt{Y}$, and $\texttt{Z}$ are the Pauli matrices.
The corresponding POVM is proposed as
\begin{equation}
    \label{eq:improved-povm}
    E^{b_i}_{i}=\frac{1}{2}\mathbb{I}^{\otimes k}+\frac{(-1)^{b_i}}{2}\xi_i .
\end{equation}
For the above (Q)RAC $\big\{\{\rho({\bf b})\},\{E^{0}_{i},E^{1}_{i}\}\big\}$, both the average and worst-case decoding success probabilities are given by
\begin{equation}
    \label{eq:improved-prob}
    \frac{1}{2}+\frac{1}{2}\sqrt{\frac{1}{(2^k-1)L}}
\end{equation}
This holds for both classical and quantum RACs.

While this value is significantly below the known upper bounds, at present few general explicit constructions are known that yields $(L,k)$-(Q)RACs for arbitrary $L$ and $k$ beyond the above approach.

In addition, the polar codes~\cite{arikan2009channel,korada2010polar} can be used to construct $(L,k)$-RACs whose average decoding success probability asymptotically attains the entropy-based upper bounds in \eqref{eq:rac-average-upper} and \eqref{eq:qrac-average-upper}.
Thus, this construction is asymptotically optimal in terms of the average decoding success probability.
However, this asymptotic optimality does not imply optimality in the non-asymptotic regime, where the optimal average decoding success probability remains unknown in general.

Very recently, a construction of an $(L,L-1)$-QRAC achieving the conjectured upper bound in Eq.~\eqref{eq:qrac-conjecture} was proposed in~\cite{suzuki2026analytical}.
However, for general $(L,k)$, an optimal construction of QRACs remains unknown, and it is still unknown whether the upper bound in \eqref{eq:qrac-conjecture} remains valid in the regimes $k=o(L)$ or $L-k=o(L)$.

In summary, the current open questions on (Q)RACs can be stated as follows:
\begin{enumerate}
    \item The actual gap between RACs and QRACs in the non-asymptotic regime with respect to both the average and worst-case decoding success probabilities.
    \item Constructive methods to build optimal $(L,k)$-(Q)RACs for arbitrary $L$ and $k$.
\end{enumerate}
The known results are summarized in Table~\ref{tab:summary-known}.

In this work, we provide partial answers to these questions.
Our main contributions are summarized as follows:
\begin{enumerate}
    \item For arbitrary $L$ and $k$, we show that constructing an average-case optimal $(L,k)$-RAC is equivalent to minimizing the directed Chamfer dissimilarity over all subsets $S\subset\{0,1\}^L$ with $|S|=2^k$ (Theorem~\ref{theorem:average1}). As a consequence, whenever a binary $t$-perfect code of length $L$ and size $2^k$ exists, it yields an average-case optimal RAC.
    \item For arbitrary $L$ and $k$, we show that constructing a worst-case optimal $(L,k)$-RAC is equivalent to minimizing the directed Hausdorff distance over all subsets of $[0,1]^L$ of size $2^k$ (Theorem~\ref{theorem:worst}).
    \item For $k=1$, we give an explicit construction of $(L,1)$-QRAC whose average decoding success probability is strictly greater that that of average-optimal $(L,1)$-RAC for all $L>1$ (Theorem~\ref{theorem:L1qrac}).
    \item For $L=2^k-1$, we prove that the binary simplex code realizes a worst-case optimal classical $(L,k)$-RAC (Theorem~\ref{theorem:worst-2k1-b}). We further give an explicit $(2^k-1,k)$-RAC, based on the states and POVMs proposed in~\cite{liabotro2017improved} (Corollary~\ref{corollary:worst-2k1}). In addition, we show that there exists a  $(2^k-1,k)$-QRAC whose worst-case decoding success probability is strictly larger than the classical optimum (Theorem~\ref{theorem:worst-2k1-Q}). This establishes a classical--quantum separation for this infinite family.
    \item For $k=L-1$, we provide an explicit classical RAC construction that is optimal under the average-case criterion and, assuming a stated conjecture, also optimal under the worst-case criterion (Theorem~\ref{theorem:upper3}). One instance of this construction is realized by the single parity-check code (Corollary~\ref{corollary:single-parity-check}). As a by-product, the same viewpoint gives a RAC-parallel derivation of known explicit $(L,L-1)$-QRACs attaining the value of an upper bound conjectured in prior work to be tight (Theorem~\ref{theorem:optimal-qrac}).
\end{enumerate}

Table~\ref{tab:summary} summarizes the optimal RACs obtained in this paper together with the corresponding optimal codes.
It also includes explicitly constructed QRACs whose decoding success probabilities are proved to be higher than those of the corresponding RACs.
Figure~\ref{fig:Lk-QRAC} plots the analytically obtained decoding success probabilities of $(7,k)$-(Q)RACs.
In previous studies, only a small number of analytical results were available in this figure.
In contrast, the present work substantially increases the number of analytically characterized points.
For $k=1,3,6$, the quantum decoding success probabilities (blue plots) lie above the corresponding classical ones (orange plots), visually demonstrating the existence of a classical--quantum gap.

In addition, we conducted numerical experiments for small values of $L$ and $k$ to investigate the gap between RACs and QRACs in terms of both average and worst-case decoding success probabilities.
Note that an explicit construction of an optimal $(L,1)$-RAC was already given in~\cite{ambainis2024quantum}; however, our proposed approach also yields a simple proof of its optimality.
Moreover, although an $(L,L-1)$-QRAC attaining the conjectured bound was already proposed in~\cite{suzuki2026analytical}, the constructions of the states and POVMs in this paper are different from those in~\cite{suzuki2026analytical}.
A novel feature of our construction is its close structural similarity to the corresponding $(L,L-1)$-RAC.
Throughout this work we restrict attention to two-level systems ($d=2$) and do not consider shared randomness or shared entanglement.


\section{Definition}
In this section, we first introduce the definition of quantum random access codes (QRACs).
We then define their classical counterpart, random access codes (RACs).
Next, we formalize two performance measures for (Q)RACs: the average decoding success probability and the worst-case decoding success probability.
Finally, we present the definition of the distance measure used in this paper.
\subsection{Definitions of (Q)RACs}
\begin{definition}[Quantum Random Access Codes]
\label{def:}
Let $L,k\in\mathbb{N}$ satisfy $1\le k\le L$, and let ${\bf b}=(b_0,\dots,b_{L-1})\in\{0,1\}^L$ be a binary string of length $L$.
For each ${\bf b}\in\{0,1\}^L$, we associate a quantum state $\rho({\bf b})$ on $k$ qubits, i.e., a density operator $\rho({\bf b})\in\mathbb{C}^{2^k\times 2^k}$ satisfying
\begin{equation}
    \left\{
    \begin{array}{l}
        \rho({\bf b}) \in \mathbb{C}^{2^k\times 2^k},\\
        \rho({\bf b})^\dagger = \rho({\bf b}),\\
        \rho({\bf b}) \succeq 0,\\
        {\rm tr}\big(\rho({\bf b})\big) = 1.
    \end{array}
    \right.
\end{equation}
The mapping ${\bf b}\mapsto \rho({\bf b})$ is referred to as the encoding of ${\bf b}$ into $k$ qubits.

Next, we consider the decoding procedure that aims to recover a bit $b_i$ from the encoded state $\rho({\bf b})$ via quantum measurement.
For each position $i\in\{0,\dots,L-1\}=:[L]$, we fix a two-outcome POVM $\{E_i^0,E_i^1\}$.
Let $B_i\in\{0,1\}$ denote the classical random variable corresponding to the measurement outcome when $\rho({\bf b})$ is measured with $\{E_i^0,E_i^1\}$.
Its distribution is given by
\begin{equation}
    \Pr(B_i=x)={\rm tr}\big(E^{x}_i\,\rho({\bf b})\big),\qquad x\in\{0,1\}.
\end{equation}
Here, the measurement operators satisfy, for all $x\in\{0,1\}$ and $i\in[L]$,
\begin{equation}
    \left\{
    \begin{array}{l}
        E^x_i \in \mathbb{C}^{2^k\times 2^k},\\
        (E^x_i)^\dagger=E^x_i,\\
        E^x_i \succeq 0,\\
        \sum_x E^x_i = \mathbb{I}^{\otimes k},
    \end{array}
    \right.
\end{equation}
where $\mathbb{I}$ denotes the identity operator on a single qubit.
The mapping $\rho({\bf b})\mapsto B_i$ is referred to as the decoding of the $i$-th bit from $k$ qubits.

A protocol consisting of such encoding and decoding is called a quantum random access code (QRAC).
In particular, we refer to it as an $(L,k)$-QRAC to emphasize the length $L$ of the original bit string and the number of message qubits $k$ used for encoding.
\end{definition}

\begin{definition}[Random Access Codes]
\label{def:rac}
A random access code (RAC) is defined as a special case of a QRAC in which both the encoding states $\{\rho({\bf b})\}$ and the decoding POVMs $\{E_i^0,E_i^1\}$ are restricted to be diagonal in the computational basis.
In particular, we call it an $(L,k)$-RAC to emphasize the length $L$ of the original bit string and the number of message (classical) bits $k$ used for the encoding.
\end{definition}

Since $\rho({\bf b})\succeq 0$ and ${\rm tr}(\rho({\bf b}))=1$, if $\rho({\bf b})$ is restricted to be diagonal in the computational basis, then its diagonal entries admit a natural probabilistic interpretation.
More precisely, letting $(\rho({\bf b}))_{mm}$ denote the $m$-th diagonal element, we may regard it as the classical conditional probability that an encoded message $M={\bf m}\in\{0,1\}^k$ is produced given the input string $B={\bf b}\in\{0,1\}^L$:
\[\displaystyle{
    P_{\rm E}(M={\bf m}\mid B={\bf b})=\big(\rho({\bf b})\big)_{mm},
}\]
\begin{equation}
    \sum_{{\bf m}\in\{0,1\}^k}P_{\rm E}(M={\bf m}\mid B={\bf b})=1.
\end{equation}
Here ${\bf m}\in\{0,1\}^k$ is the $k$-bit binary representation of $m\in[2^k]$.
Thus, in RACs, the encoding procedure corresponds to stochastically generating a $k$-bit message from an $L$-bit input string.

Similarly, if the POVM $\{E^0_i,E^1_i\}$ is restricted to be diagonal in the computational basis, then its $m$-th diagonal element can be interpreted as the classical conditional probability of outputting a bit value $B'_i=b'_i\in\{0,1\}$ given the message $M={\bf m}$:
\[\displaystyle{
    P_{\rm D}(B'_i=b'_i\mid M={\bf m})=\big(E^{b'_i}_i\big)_{mm},
}\]
\begin{equation}
    \sum_{b'_i\in\{0,1\}}P_{\rm D}(B'_i=b'_i\mid M={\bf m})=1.
\end{equation}
Indeed, computing ${\rm tr}\big(E^{b'_i}_{i}\rho({\bf b})\big)$ yields
\begin{align}
    {\rm tr}\big(E^{b'_i}_{i}\rho({\bf b})\big)
    &=\sum_{m\in[2^k]}\big(E^{b'_i}_{i}\big)_{mm}\big(\rho({\bf b})\big)_{mm}\notag\\
    &=\sum_{{\bf m}\in\{0,1\}^k}P_{\rm D}(B'_i=b'_i\mid M={\bf m})\notag\\
    &\hspace{10mm}\cdot P_{\rm E}(M={\bf m}\mid B={\bf b})\notag\\
    &=\Pr(B'_i=b'_i\mid B={\bf b}),
\end{align}
which is precisely the probability that the decoding procedure applied to ${\bf b}\in\{0,1\}^L$ outputs the bit value $b'_i\in\{0,1\}$.
In what follows, whenever it is clear from the context, we write $P_{\rm E}(M={\bf m}\mid B={\bf b})$ as $P_{\rm E}({\bf m}\mid {\bf b})$ and $P_{\rm D}(B'_i=b'_i\mid M={\bf m})$ as $P_{\rm D}(b'_i\mid {\bf m})$ for simplicity.

\subsection{Average and worst-case decoding success probabilities}
\label{sec:decoding-success}
In what follows, we say that decoding is successful whenever $b'_i=b_i$ holds.

\begin{definition}[Average and worst-case decoding success probabilities]
\label{def:avg-worst}
The average decoding success probability of an $(L,k)$-QRAC is defined as
\begin{equation}
    \label{eq:q-average}
    \frac{1}{L\cdot 2^L}\sum_{i\in[L],{\bf b}\in\{0,1\}^L}{\rm tr}\big(E^{b_i}_i\rho({\bf b})\big),
\end{equation}
whereas the worst-case decoding success probability of an $(L,k)$-QRAC is defined as
\begin{equation}
    \label{eq:q-worst}
    \min_{i\in[L],\,{\bf b}\in\{0,1\}^L}{\rm tr}\big(E^{b_i}_i\rho({\bf b})\big).
\end{equation}
Likewise, the average decoding success probability of an $(L,k)$-RAC is defined as
\begin{equation}
    \label{eq:def-average}
    \frac{1}{L\cdot2^L}\sum_{i\in[L],{\bf b}\in\{0,1\}^L}\Pr(b_i\mid {\bf b}),
\end{equation}
whereas the worst-case decoding success probability of an $(L,k)$-RAC is defined as
\begin{equation}
    \label{eq:worst}
    \min_{i\in[L],\,{\bf b}\in\{0,1\}^L}\Pr(b_i\mid {\bf b}).
\end{equation}
\end{definition}

\subsection{Distances}
\label{sec:distances}
We introduce below the fundamental distance measures used in this paper.
\begin{definition}
The Hamming distance $d_{\rm H}$ and the relative Hamming distance $d_{{\rm H}/L}$ between ${\bf a}, {\bf b} \in\{0,1\}^L$ are defined as follows:
\begin{align}
    d_{\rm H}({\bf a},{\bf b})&:=\sum_{i\in[L]}{\bf 1}_{a_i\neq b_i},\\
    d_{{\rm H}/L}({\bf a},{\bf b})&:=\frac{1}{L}d_{\rm H}({\bf a},{\bf b}).
\end{align}
In addition, the Chebyshev distance $d_\infty$ between ${\bf a}, {\bf b} \in\mathbb{R}^L$ is defined as follows:
\begin{align}
    d_\infty({\bf a},{\bf b})&:=\max_{i\in[L]}|a_i-b_i|.
\end{align}
\end{definition}

\section{Optimal random access codes}
\label{sec:rac}

In this section, we first reformulate the problems of finding average-optimal and worst-case optimal $(L,k)$-RACs, originally defined in terms of maximizing decoding probabilities, as geometric optimization problems of finding point sets that minimize certain distance-like measures.
We then derive explicit constructions of average-optimal and worst-case optimal RACs for several parameter families $(L,k)$ and show that these constructions can be achieved by well-known infinite family of linear codes in coding theory.
Furthermore, for some parameter families, we present QRAC constructions whose decoding success probabilities are provably strictly higher than those of the corresponding RACs, thereby demonstrating the existence of a classical--quantum gap.

\subsection{Average-Optimal \texorpdfstring{$(L,k)$}{(L,k)}-RAC}
\label{sec:rac-average-upper}
We first formulate the problem of constructing average-optimal $(L,k)$-RACs as follows:
\begin{theoremE}[][normal]
\label{theorem:average1}
Let $k,L\in\mathbb{N}$ satisfy $1\le k\le L$.
The maximum achievable average decoding success probability of an $(L,k)$-RAC is 
\begin{equation}
\label{eq:average-prob}
    1- \min_{\substack{S\subset\V\\|S|=2^k}} d_{\mathrm{Cham}}^{\rightarrow}\big(\V,S;\,d_{{\rm H}/L}\big)
\end{equation}
where
\begin{equation}
    d_{\mathrm{Cham}}^{\rightarrow}(A,B;d):=\frac{1}{|A|}\sum_{{\bf a}\in A}\inf_{{\bf b}\in B}d({\bf a},{\bf b}).
\end{equation}
is the directed Chamfer dissimilarity~\cite{barrow1977parametric}, also known as the directed modified Hausdorff distance~\cite{dubuisson1994modified}.
\end{theoremE}
\begin{proofE}
By definition, the average decoding error probability can be written as
\begin{align}
     \epsilon_{\rm avg}(P_{\rm E},P_{\rm D})
     & = \frac{1}{L\cdot2^L}\sum_{{\bf b},i}\Pr(B'_i\neq b_i\mid {\bf b})\notag\\
     & = \frac{1}{2^L}\sum_{{\bf b},{\bf m},i}\frac{1}{L}P_{\rm D}(B'_i\neq b_i\mid {\bf m})\,P_{\rm E}({\bf m}\mid {\bf b})\notag\\
     & = \frac{1}{2^L} \sum_{{\bf b},{\bf b}',{\bf m}}d_{{\rm H}/L}({\bf b},{\bf b}')\,P_{\rm D}({\bf b}'\mid {\bf m})\,P_{\rm E}({\bf m}\mid {\bf b})\notag\\
     & = \frac{1}{2^L} \sum_{{\bf b}\in \V}\Delta({\bf b},P_{\rm E},P_{\rm D}). \label{eq:eps-avg}
\end{align}
Here, $P_{\rm D}(B'={\bf b}'\mid M={\bf m})$ denotes a conditional joint distribution whose marginal distributions satisfy $P_{\rm D}(B'_i=b'_i\mid M={\bf m}) $ for each $i$.
In addition, we used the identity
\begin{align}
    &\frac{1}{L}\sum_{i\in[L]}P_{\rm D}(B'_i\neq b_i\mid {\bf m})\notag\\
    &\qquad=\frac{1}{L}\sum_{i\in[L]}\sum_{\substack{{\bf b}'\in \V\\b'_i \neq b_i}}P_{\rm D}({\bf b}'\mid {\bf m})\notag\\
    &\qquad=\frac{1}{L}\sum_{i,{\bf b}'}{\bf 1}_{b'_i\neq b_i}\,P_{\rm D}({\bf b}'\mid {\bf m})\notag\\
    &\qquad=\sum_{{\bf b}'}d_{{\rm H}/L}({\bf b},{\bf b}')\,P_{\rm D}({\bf b}'\mid {\bf m}), \label{eq:avg-bit-to-hamming}
\end{align}
and defined
\begin{equation}
\label{eq:Delta-def}
     \Delta({\bf b},P_{\rm E},P_{\rm D})
     := \sum_{{\bf m},{\bf b}'}d_{{\rm H}/L}({\bf b},{\bf b}')\,P_{\rm D}({\bf b}'\mid {\bf m})\,P_{\rm E}({\bf m}\mid {\bf b}).
\end{equation}

Define
\begin{equation}
\label{eq:FD-def}
\mathcal{F}_{\rm D}({\bf b},{\bf m};P_{\rm D})
:=\sum_{{\bf b}'}d_{{\rm H}/L}({\bf b},{\bf b}')\,P_{\rm D}({\bf b}'\mid {\bf m}),
\end{equation}
and
\begin{equation}
\label{eq:FE-def}
\mathcal{F}_{\rm E}({\bf b}',{\bf m};P_{\rm E})
:=\sum_{{\bf b}}d_{{\rm H}/L}({\bf b},{\bf b}')\,P_{\rm E}({\bf m}\mid {\bf b}).
\end{equation}
Then $\epsilon_{\rm avg}$ can be expressed as
\begin{equation}
\label{eq:e-avg-d}
     \epsilon_{\rm avg}(P_{\rm E},P_{\rm D})
     = \frac{1}{2^L} \sum_{{\bf m},{\bf b}}\mathcal{F}_{\rm D}({\bf b},{\bf m};P_{\rm D})\,P_{\rm E}({\bf m}\mid {\bf b}),
\end{equation}
and also as
\begin{equation}
\label{eq:e-avg-e}
     \epsilon_{\rm avg}(P_{\rm E},P_{\rm D})
     = \frac{1}{2^L} \sum_{{\bf m},{\bf b}'}\mathcal{F}_{\rm E}({\bf b}',{\bf m};P_{\rm E})\,P_{\rm D}({\bf b}'\mid {\bf m}).
\end{equation}
Since the right-hand side of \eqref{eq:e-avg-d} (resp.\ \eqref{eq:e-avg-e}) is a convex combination of $\mathcal{F}_{\rm D}({\bf b},{\bf m};P_{\rm D})$ (resp.\ $\mathcal{F}_{\rm E}({\bf b}',{\bf m};P_{\rm E})$), it suffices to consider deterministic encoders and deterministic decoders.
This coincides with that of \cite[Lemma~1]{saha2023measurement}, namely, that an average-optimal RAC can be realized using deterministic encoder and decoder.

We now show the following two facts, which allow us to rewrite the minimization over $(P_{\rm E},P_{\rm D})$ as a minimization over $(S,f)$:
\begin{itemize}
    \item For any deterministic encoder--decoder pair $(P_{\rm E}^{\rm det},P_{\rm D}^{\rm det})$, there exist a set $S\subset\{0,1\}^L$ with $|S|\le 2^k$ and a map $f:\{0,1\}^L\rightarrow S$ such that
    $\Delta({\bf b},P^{\rm det}_{\rm E},P^{\rm det}_{\rm D})=d_{{\rm H}/L}\big({\bf b},f({\bf b})\big)$.
    \item Conversely, for any $S\subset\{0,1\}^L$ with $|S|\le 2^k$ and any map $f:\{0,1\}^L\rightarrow S$, there exists a deterministic encoder--decoder pair $(P_{\rm E}^{\rm det},P_{\rm D}^{\rm det})$ such that
    $d_{{\rm H}/L}\big({\bf b},f({\bf b})\big)=\Delta({\bf b},P^{\rm det}_{\rm E},P^{\rm det}_{\rm D})$.
\end{itemize}

First, fix an arbitrary deterministic decoder $P_{\rm D}^{\rm det}$ and define, for each message ${\bf m}\in\{0,1\}^k$, the decoded string by
\begin{equation}
\label{eq:b-of-m}
{\bf b}^{({\bf m})}(P^{\rm det}_{\rm D})
:=\sum_{{\bf b}' \in \{0,1\}^L} {\bf b}'\, P^{\rm det}_{\rm D}({\bf b}'\mid {\bf m}).
\end{equation}
Since $P^{\rm det}_{\rm D}$ is deterministic and ${\bf b}'\in \V$, we have ${\bf b}^{({\bf m})}(P^{\rm det}_{\rm D})\in \V$.
Collecting these decoded strings over all messages yields
\begin{equation}
    \label{eq:s-det}
    S(P_{\rm D}^{\rm det})
    :=\big\{{\bf b}^{({\bf m})}(P^{\rm det}_{\rm D}) \bigm| {\bf m}\in\{0,1\}^k\big\}\subset \V.
\end{equation}
Because $|\{0,1\}^k|=2^k$ and duplicates are allowed, we have $|S(P_{\rm D}^{\rm det})|\le 2^k$.

Now suppose a deterministic encoder maps each ${\bf b}\in \V$ to a unique message ${\bf m}^{({\bf b})}\in\{0,1\}^k$, i.e.,
\begin{equation}
\label{eq:det-encoder}
P^{\rm det}_{\rm E}({\bf m}\mid {\bf b})=
\left\{
\begin{array}{ll}
     1 & {\bf m}={\bf m}^{({\bf b})},\\
     0 & \text{otherwise}.
\end{array}
\right.
\end{equation}
Then
\begin{align}
\Delta({\bf b},P^{\rm det}_{\rm E},P^{\rm det}_{\rm D})
&=\sum_{{\bf m}\in\{0,1\}^k}d_{{\rm H}/L}({\bf b},{\bf b}^{({\bf m})})\,P^{\rm det}_{\rm E}({\bf m}\mid {\bf b})\notag\\
&=d_{{\rm H}/L}\big({\bf b},f({\bf b})\big),\label{eq:f-2}
\end{align}
where we defined
\begin{equation}
    \label{eq:f-det}
    f({\bf b})
    :={\bf b}^{({\bf m}^{({\bf b})})}.
\end{equation}
By \eqref{eq:s-det}, ${\bf b}^{({\bf m})}\in S(P_{\rm D}^{\rm det})$ for all ${\bf m}$, and hence $f({\bf b})\in S(P_{\rm D}^{\rm det})$. This proves the first item.

Next, suppose we are given a set
\begin{equation}
\label{eq:S-given}
S=\big\{{\bf b}^{(1)},{\bf b}^{(2)},\dots\big\}\subset\{0,1\}^L,
\quad
|S|\le 2^k,
\end{equation}
and a map
\begin{equation}
\label{eq:f-given}
f:\{0,1\}^L\rightarrow S.
\end{equation}
We construct a deterministic decoder by assigning, for each message ${\bf m}\in\{0,1\}^k$, a fixed output ${\bf b}^{({\bf m})}\in S$:
\begin{equation}
\label{eq:pd-2}
P_{\rm D}^{\rm det}({\bf b}'\mid {\bf m})=
\left\{
\begin{array}{ll}
     1 & {\bf b}'={\bf b}^{({\bf m})},\\
     0 & \text{otherwise}.
\end{array}
\right.
\end{equation}
Then
\begin{equation}
    \sum_{{\bf b}'\in\V}d_{{\rm H}/L}({\bf b},{\bf b}')\,P_{\rm D}^{\rm det}({\bf b}'\mid {\bf m})
    = d_{{\rm H}/L}({\bf b},{\bf b}^{({\bf m})}).
\end{equation}
Since $f({\bf b})\in S$, we can choose the indexing so that for each ${\bf b}$ we have
\begin{equation}
    f({\bf b})={\bf b}^{({\bf m}^{({\bf b})})}
\end{equation}
for some message ${\bf m}^{({\bf b})}$. Define a deterministic encoder by
\begin{equation}
\label{eq:pe-2}
    P_{\rm E}^{\rm det}({\bf m}\mid {\bf b})=
    \left\{
    \begin{array}{ll}
    1 & {\bf m}={\bf m}^{({\bf b})},\\
    0 & \text{otherwise}.
    \end{array}
    \right.
    \end{equation}
Then
\begin{align}
    \Delta({\bf b},P_{\rm E}^{\rm det},P_{\rm D}^{\rm det})
    &=\sum_{{\bf m}\in\{0,1\}^k}d_{{\rm H}/L}({\bf b},{\bf b}^{({\bf m})})\,P_{\rm E}^{\rm det}({\bf m}\mid {\bf b})\notag\\
    &=d_{{\rm H}/L}\big({\bf b},f({\bf b})\big),
\end{align}
which establishes the second item.

Combining the two items, minimizing over deterministic encoder--decoder pairs is equivalent to minimizing over $(S,f)$:
\[\displaystyle{
\min_{P_{\rm E},P_{\rm D}}\sum_{{\bf b}\in \V}\Delta({\bf b},P_{\rm E},P_{\rm D})
}\]
\begin{equation}
=\min_{\substack{S\subset \V\\|S|\le 2^k}}\sum_{{\bf b}\in \V}\min_{{\bf s}\in S}d_{{\rm H}/L}\big({\bf b},{\bf s}\big).
\end{equation}
Therefore,
\begin{align}
    \min_{P_{\rm E},P_{\rm D}}\epsilon_{\rm avg}(P_{\rm E},P_{\rm D})
    &=\min_{\substack{S\subset \V\\|S|\le 2^k}}\frac{1}{2^L}\sum_{{\bf b}\in \V}\min_{{\bf s}\in S}d_{{\rm H}/L}({\bf b},{\bf s})\notag\\
    &=\min_{\substack{S\subset \V\\|S|\le 2^k}}d_{\mathrm{Cham}}^{\rightarrow}\big(\V,S;\,d_{{\rm H}/L}\big)\label{eq:eps-avg-min-to-delta}
\end{align}
where
\begin{equation}
    d_{\mathrm{Cham}}^{\rightarrow}(A,B;d):=\frac{1}{|A|}\sum_{{\bf a}\in A}\inf_{{\bf b}\in B}d({\bf a},{\bf b})
\end{equation}
is a directed Chamfer dissimilarity.
Moreover, if $|S|<2^k$, we can add arbitrary points to $S$ without increasing the objective value, and hence we may restrict to $|S|=2^k$:
\begin{equation}
    \label{eq:eps-avg-min-to-delta-eq}
    \min_{P_{\rm E},P_{\rm D}}\epsilon_{\rm avg}(P_{\rm E},P_{\rm D})
    =\min_{\substack{S\subset \V\\|S|= 2^k}}d_{\mathrm{Cham}}^{\rightarrow}\big(\V,S;\,d_{{\rm H}/L}\big).
\end{equation}
Thus the maximum average decoding success probability of an $(L,k)$-RAC is
\begin{align}
    &1-\min_{P_{\rm E},P_{\rm D}}\epsilon_{\rm avg}(P_{\rm E},P_{\rm D})\notag\\
    &=1-\min_{\substack{S\subset \V\\|S|= 2^k}}d_{\mathrm{Cham}}^{\rightarrow}\big(\V,S;\,d_{{\rm H}/L}\big).
\end{align}
\end{proofE}

\subsubsection{Constructing average-optimal \texorpdfstring{$(L,k)$}{(L,k)}-RAC}
\label{sec:rac-average-construct}
To construct an average-optimal $(L,k)$-RAC, one needs to solve the optimization problem 
\begin{equation}
    \label{eq:prob-average}
    \min_{\substack{S\subset\V\\|S|=2^k}} 
    \frac{1}{2^L}\sum_{{\bf b}\in \V}\min_{{\bf s}\in S}d_{{\rm H}/L}({\bf b},{\bf s})
\end{equation}
This problem can be formulated as a mixed-integer linear program (MILP) as follows.

Introduce binary variables $z_{\bf a}\in\{0,1\}$ indicating whether ${\bf a}\in \V$ is selected into $S$ (i.e., $z_{\bf a}=1$ if ${\bf a}\in S$, and $z_{\bf a}=0$ otherwise).
Also introduce binary assignment variables $\lambda_{{\bf a}}({\bf b})\in\{0,1\}$ indicating whether ${\bf b}\in \V$ is assigned to ${\bf a}\in \V$ (i.e., $\lambda_{\bf a}({\bf b})=1$ if ${\bf a}$ is chosen as the representative for ${\bf b}$, and $\lambda_{\bf a}({\bf b})=0$ otherwise).
The objective is to minimize the total (relative) Hamming distance between each ${\bf b}$ and its assigned representative ${\bf a}$.
After optimization, for each ${\bf b}$ the variable $\lambda_{\bf a}({\bf b})$ identifies a nearest representative in the selected set $S$.
The resulting MILP is
\begin{align}
    & \text{minimize} && \sum_{{\bf b}\in \V}\sum_{{\bf a}\in \V}d_{\rm H}({\bf b},{\bf a})\,\lambda_{\bf a}({\bf b}) \notag\\
    & \text{subject to} && \sum_{{\bf a}\in \V}z_{\bf a}=2^k,\notag\\
    & \phantom{\text{subject to}} && \sum_{{\bf a}\in \V}\lambda_{\bf a}({\bf b})=1 &&\hspace{-20mm} \forall {\bf b}\in \V,\notag\\
    & \phantom{\text{subject to}} && z_{\bf a}\in\{0,1\} && \hspace{-20mm}\forall {\bf a}\in \V,\notag\\
    & \phantom{\text{subject to}} && \lambda_{\bf a}({\bf b})\in\{0,1\} && \hspace{-20mm}\forall {\bf a}, {\bf b}\in \V,\notag\\
    & \phantom{\text{subject to}} && \lambda_{\bf a}({\bf b})\le z_{\bf a} && \hspace{-20mm}\forall {\bf a}, {\bf b}\in \V.\label{eq:milp-average}
\end{align}

After solving \eqref{eq:milp-average}, one can define an encoder and a decoder as follows.
Fix a bijection between the selected representatives in $S$ and the message set $\{0,1\}^k$, and denote the corresponding message for ${\bf s}^{(j)}\in S$ by ${\bf m}^{(j)}$.
Then define the encoder by assigning ${\bf b}$ to the message corresponding to its selected representative:
\begin{equation}
    P_{\rm E}({\bf m}^{(j)}\mid {\bf b})=\lambda_{{\bf s}^{(j)}}({\bf b}).
\end{equation}
Define the decoder deterministically by
\begin{equation}
    P_{\rm D}({\bf b}'\mid {\bf m}^{(j)})=\left\{
    \begin{array}{ll}
        1 & {\bf b}'={\bf s}^{(j)},\\
        0 & \text{otherwise}.
    \end{array}
    \right.
\end{equation}
In this formulation, for each ${\bf b}$ we allow only a single nearest representative in $S$; however, if there are multiple representatives at the same minimum Hamming distance, any tie-breaking rule yields the same average success probability.

Alternatively, one may directly construct an encoder and a decoder from a given set $S$.
Although finding an $S$ that maximizes the average success probability takes an exponentially large time in general, if an optimal $S$ is obtained by some method, then one can build an average-optimal $(L,k)$-RAC as follows.
First, define the decoder by assigning each message ${\bf m}^{(j)}$ to a distinct ${\bf b}^{*(j)}\in S$:
\begin{equation}
    P_{\rm D}({\bf b}^{*(j)}\mid {\bf m}^{(j)})=\left\{
    \begin{array}{ll}
        1 & D({\bf m}^{(j)})={\bf b}^{*(j)},\\
        0 & \text{otherwise}.
    \end{array}
    \right.
\end{equation}
That is, the $j$-th message ${\bf m}^{(j)}$ is decoded to the $j$-th element ${\bf b}^{*(j)}$ of $S$.
Next, define the encoder by choosing a message that decodes to the nearest codeword in $S$:
\[\displaystyle{
    P_{\rm E}({\bf m}^{(j)}\mid {\bf b})
}\]
\begin{equation}
    \label{eq:ml-decode-1}
    =\left\{
    \begin{array}{ll}
        >0 & {\bf m}^{(j)}\in\big\{D^{-1}\big(\big\{{\bf b}'\bigm| d_{\rm H}({\bf b},{\bf b}')=d_{\rm H}^*({\bf b})\big\}\big)\big\},\\
        0 & \text{otherwise},
    \end{array}
    \right.
\end{equation}
where
\begin{equation}
    \label{eq:ml-decode-2}
    d_{\rm H}^*({\bf b})=\min_{{\bf b}'\in S}d_{\rm H}({\bf b},{\bf b}').
\end{equation}
In other words, the encoder selects a message whose decoded ${\bf b}'\in S$ is at minimum Hamming distance from ${\bf b}$.
If there are multiple nearest bit strings in $S$, one may either distribute probability among the corresponding messages, or select any one of them deterministically; both choices yield the same average success probability.

\subsubsection{Closed-form upper bound of average decoding success probability}
Theorem~\ref{theorem:average1} shows that an average-optimal $(L,k)$-RAC can be obtained by solving an MILP.
However, this problem generally requires exponential time, and the formulation is not well suited for proving optimality for specific families of $(L,k)$.
Therefore, we derive the following closed-form upper bound, which is more convenient for theoretical analysis.

\begin{theoremE}[][normal]
\label{theorem:average}
Let
\begin{equation}
H=\min\Big\{h\in\mathbb{N}\Bigm|\sum^h_{l=0}\binom{L}{l}\ge2^{L-k}\Big\}.
\end{equation}
Then the average decoding success probability of an $(L,k)$-RAC is upper bounded by
\begin{equation}
\label{eq:average}
    1-\frac{1}{L\cdot 2^{L-k}}\Bigg(\sum^{H-1}_{h=0}h\binom{L}{h}
    +H\Big(2^{L-k}-\sum^{H-1}_{h'=0}\binom{L}{h'}\Big)\Bigg).
\end{equation}
\end{theoremE}
\begin{proofE}
For a fixed ${\bf b}^{*(i)}\in S$, the number of bit strings at Hamming distance exactly $l$ from ${\bf b}^{*(i)}$ is at most $\binom{L}{l}$.
Hence, the total number of bit strings within Hamming distance at most $h$ from the $2^k$ elements in $S$ is at most
\begin{equation}
    N(h)=2^k\sum^{h}_{l=0}\binom{L}{l}.
\end{equation}
Choose the smallest $h\in\mathbb{N}$ such that $N(h)\ge 2^L$, and denote it by $H$. Equivalently,
\begin{equation}
    H=\min\Big(h\in\mathbb{N}\Bigm|\sum^h_{l=0}\binom{L}{l}\ge2^{L-k}\Big).
\end{equation}
Let $S^{(h)}$ denote the set of bit strings at minimum Hamming distance $h$ from $S$, i.e.,
\begin{equation}
    \label{eq:Sh}
    S^{(h)}=\big\{{\bf b}\bigm|\min_{{\bf b}^*\in S}d_{\rm H}({\bf b}, {\bf b}^*)=h\big\}.
\end{equation}
We have $|S^{(h)}|\le 2^k\binom{L}{h}$, and in the ideal (non-overlapping) case one may think of $|S^{(h)}|=2^k\binom{L}{h}$.
To maximize the average success probability, we would like to make the distances $\min_{{\bf s}\in S}d_{{\rm H}/L}({\bf b},{\bf s})$ as small as possible for as many ${\bf b}\in \V$ as possible.
If we assume that the layers $S^{(h)}$ can be arranged without overlaps up to radius $H$, that is, there is no ${\bf b}\in\V$ such that $d_{\rm H}({\bf b},{\bf b}^*)=d_{\rm H}({\bf b},{{\bf b}^*}')\le H$ for ${\bf b}^*,{{\bf b}^*}'\in S$ and ${\bf b}^*\neq {{\bf b}^*}'$, then the sum of normalized distances is lower bounded by
\begin{equation}
    \frac{1}{L}\Big(\sum^{H-1}_{h=0}h\cdot\big|S^{(h)}\big|+H\cdot\Big(2^L-\sum^{H-1}_{h=0}\big|S^{(h)}\big|\Big)\Big).
\end{equation}
Since such an ideal partition may not exist in general, this yields only a lower bound on the average error probability. Concretely,
\[\displaystyle{
    \min_{P_{\rm E},P_{\rm D}}\epsilon_{\rm avg}(P_{\rm E},P_{\rm D})
}\]
\begin{equation}
    \ge \frac{1}{L\cdot 2^{L-k}}
    \Bigg(\sum^{H-1}_{h=0}h\binom{L}{h}+H\Big(2^{L-k}-\sum^{H-1}_{h'=0}\binom{L}{h'}\Big)\Bigg).
\end{equation}
Therefore, the average decoding success probability satisfies
\[\displaystyle{
1-\epsilon_{\rm avg}
}\]
\begin{equation}
\le
1-\frac{1}{L\cdot 2^{L-k}}
\Bigg(\sum^{H-1}_{h=0}h\binom{L}{h}+H\Big(2^{L-k}-\sum^{H-1}_{h'=0}\binom{L}{h'}\Big)\Bigg),
\end{equation}
which proves \eqref{eq:average}.
\end{proofE}

As is immediate from the derivation of the above closed-form upper bound, the following property holds for average-optimal $(L,k)$-RACs.

\begin{corollaryE}[][normal]
\label{corollary:perfect-t}
Binary $t$-perfect codes~\cite{macwilliams1977theory} attain the upper bound in Theorem~\ref{theorem:average}.
\end{corollaryE}
\begin{proofE}
A binary $t$-perfect code $S$ is a code such that every
$\mathbf b\in\{0,1\}^L$ lies within Hamming distance at most $t$ from exactly one codeword $\mathbf s\in S$. This is precisely the condition assumed in the proof of Theorem~\ref{theorem:average} that the layers $S^{(t)}$ do not overlap. Therefore, binary $t$-perfect codes attain the upper bound in Theorem~\ref{theorem:average}.
\end{proofE}

Note that the converse does not hold; that is, not every average-optimal $(L,k)$-RAC corresponds to a binary $t$-perfect code.

A well-known example of a binary $1$-perfect code is the binary Hamming code, for which the following holds.

\begin{corollaryE}[][normal]
\label{corollary:hamming-code}
The average decoding success probability of an $(L,k)$-RAC for $L=2^m-1$ and $k=2^m-m-1$ is upper bounded by
\begin{equation}
    1-\frac{1}{L+1}.
\end{equation}
This upper bound is achieved by taking $S$ as binary Hamming code.
\end{corollaryE}
\begin{proofE}
For ${\bf b}\in S$, the minimum Hamming distance to a codeword ${\bf s}\in S$ is clearly zero.  
For the remaining $2^L-|S|$ points, the minimum Hamming distance is at least $1$. Hence,
\begin{equation}
    \sum_{{\bf b}\in\{0,1\}^L}\min_{{\bf s}\in S}d_{\rm H}({\bf b},{\bf s})
    \ge 2^L-2^k
    = L2^k,
\end{equation}
where we used $L=2^m-1$ and $k=2^m-m-1$.
Since the binary Hamming code is a perfect code of radius $1$, every ${\bf b}\notin S$ is at Hamming distance exactly $1$ from a unique codeword ${\bf s}\in S$.
Therefore, the above lower bound is attained.
Consequently, if $S$ is chosen to be the binary Hamming code, then
\begin{equation}
    d^\rightarrow_{\rm Cham}\big(\{0,1\}^L,S;d_{\rm H/L}\big)
    =2^{k-L}
    =\frac{1}{L+1},
\end{equation}
which is optimal.
The corresponding average decoding success probability is
\begin{equation}
    1-\frac{1}{L+1}
\end{equation}
where $L=2^m-1$.
\end{proofE}

In addition to the case $(L,k)=(2^m-1,2^m-m-1)$, the upper bound of the average decoding success probability of $(L,1)$-RAC and $(L,L-1)$-RAC can be obtained straightforwardly as follows:

\begin{corollaryE}[\cite{ambainis2008quantum}][normal]
\label{corollary:average-l1}
The average decoding success probability of an $(L,1)$-RAC is upper bounded by
\begin{equation}
    \label{eq:L1rac}
    \frac{1}{2}+\frac{1}{2^L}\binom{L-1}{\lfloor\frac{L}{2}\rfloor}.
\end{equation}
\end{corollaryE}
\begin{proofE}
By the binomial theorem and the symmetry of binomial coefficients, it can be shown that
\[
H=\left\lfloor \frac{L}{2}\right\rfloor
\]
when $k=1$, regardless of whether $L$ is even ($L=2n$ with $n\in\mathbb{N}$) or odd $(L=2n+1)$.
Moreover, using the identity
\begin{equation}
    h\binom{L}{h}=L\binom{L-1}{h-1},
\end{equation}
together with the change of variables $j=h-1$, we obtain
\begin{equation}
    \sum_{h=0}^{n-1}h\binom{L}{h}
    =
    L\sum_{j=0}^{n-2}\binom{L-1}{j}.
\end{equation}
Applying this identity, when $L=2n+1$, we have
\begin{align}
    &\sum_{h=0}^{n-1}h\binom{2n+1}{h}
    +n\left(2^{2n}-\sum_{h=0}^{n-1}\binom{2n+1}{h}\right)\notag\\
    &=L\cdot2^{L-1}\left(\frac{1}{2}-\frac{1}{2^L}\binom{L-1}{\left\lfloor\frac{L}{2}\right\rfloor}\right),
\end{align}
whereas, when $L=2n$, we obtain
\begin{align}
    &\sum_{h=0}^{n-1}h\binom{2n}{h}
    +n\left(2^{2n-1}-\sum_{h=0}^{n-1}\binom{2n}{h}\right)\notag\\
    &=L\cdot2^{L-1}\left(\frac{1}{2}-\frac{1}{2^L}\binom{L-1}{\frac{L}{2}}\right).
\end{align}
Therefore, the upper bound of the average decoding success probability of $(L,1)$-RAC is
\begin{equation}
    \frac{1}{2}+\frac{1}{2^L}\binom{L-1}{\left\lfloor\frac{L}{2}\right\rfloor}.
\end{equation}
\end{proofE}

\begin{corollaryE}[][normal]
\label{corollary:repetition-code}
The upper bound of Eq.~\eqref{eq:L1rac} is achieved by taking $S$ as repetition code.
\end{corollaryE}
\begin{proofE}
The repetition code~\cite{macwilliams1977theory} is defined by
\begin{equation}
    S=\{0^L,1^L\}.
\end{equation}
For ${\bf b}\in\{0,1\}^L$ of Hamming weight $w$, the minimum Hamming distance to a codeword ${\bf s}\in S$ is
\begin{equation}
    \min_{{\bf s}\in S} d_{\rm H}({\bf b},{\bf s})
    = \min(w,L-w).
\end{equation}
Therefore,
\begin{equation}
    \frac{1}{2^L}
    \sum_{{\bf b}\in\{0,1\}^L}
    \min_{{\bf s}\in S} d_{\rm H/L}({\bf b},{\bf s})
    =
    \frac{1}{2}
    -
    \frac{1}{2^L}
    \binom{L-1}{\left\lfloor \frac{L}{2} \right\rfloor}.
\end{equation}
Hence, the corresponding average decoding success probability is
\begin{equation}
    \frac{1}{2}
    +
    \frac{1}{2^L}
    \binom{L-1}{\left\lfloor \frac{L}{2} \right\rfloor},
\end{equation}
which coincides with the upper bound in~\eqref{eq:L1rac}.
Therefore, the repetition code is optimal.
Note that when $L$ is odd, the repetition code can be regarded as a binary $\frac{L-1}{2}$-perfect code. Therefore, this observation is consistent with Corollary~\ref{corollary:perfect-t}.
\end{proofE}

Ambainis et al.~\cite{ambainis2024quantum} showed that $(L,1)$-QRACs can outperform average-optimal $(L,1)$-RACs in average decoding success probability. 
However, this quantum advantage was rigorously verified only for relatively small values of $L$ $(L \leq 11)$.
Whether it persists for general $L$ has remained an open question.
Here, we provide an explicit construction of $(L,1)$-QRACs that achieve a strictly higher average decoding success probability than average-optimal $(L,1)$-RACs.

\begin{theoremE}[][normal]
\label{theorem:L1qrac}
There exists $(L,1)$-QRAC whose average decoding success probability is strictly greater than that of average optimal $(L,1)$-RAC for $L>1$.
\end{theoremE}
\begin{proofE}
For $L=1$, both a $(1,1)$-RAC and a $(1,1)$-QRAC achieve decoding success probability one.
Therefore, in the following we only consider the case $L>1$.

Let ${\bf r}({\bf b})$ denote the Bloch vector of the state $\rho({\bf b})$.
For a single-qubit state, we write
\begin{equation}
    \rho({\bf b})
    =
    \frac{1}{2}
    \big(
        \mathbb{I}
        +{\bf r}({\bf b})\cdot{\boldsymbol\sigma}
    \big),
    \qquad
    {\boldsymbol\sigma}:=(\texttt{X},\texttt{Y},\texttt{Z}),
\end{equation}
where $\|{\bf r}({\bf b})\|\le 1$.
For each $i\in[L]$, the optimal Helstrom measurement for discriminating the two averaged states corresponding to $b_i=0$ and $b_i=1$ gives
\begin{align}
    P_i
    &=
    \frac{1}{2}
    +
    \frac{1}{4}
    \left\|
        \frac{1}{2^{L-1}}
        \sum_{{\bf b}\,:\, b_i=0}{\bf r}({\bf b})
        -
        \frac{1}{2^{L-1}}
        \sum_{{\bf b}\,:\, b_i=1}{\bf r}({\bf b})
    \right\| \notag \\
    &=
    \frac{1}{2}
    +
    \frac{1}{2}\|{\boldsymbol\mu}_i\|,
\end{align}
where
\begin{equation}
    {\boldsymbol\mu}_i
    :=
    \frac{1}{2^L}
    \sum_{{\bf b}\in\{0,1\}^L}
    (-1)^{b_i}{\bf r}({\bf b})
    =
    \mathbb{E}_{\bf b}
    \big[
        (-1)^{b_i}{\bf r}({\bf b})
    \big],
\end{equation}
and ${\bf b}$ is uniformly distributed over $\{0,1\}^L$.
Thus, the average decoding success probability under the Helstrom measurements is
\begin{equation}
    \label{eq:L1-1}
    P
    =
    \frac{1}{2}
    +
    \frac{1}{2L}
    \sum_{i\in[L]}\|{\boldsymbol\mu}_i\|.
\end{equation}

In the present $(L,1)$-(Q)RAC model, we choose the Bloch vector as
\begin{equation}
    {\bf r}({\bf b})
    =
    \begin{cases}
        {\bf t}({\bf b})/\|{\bf t}({\bf b})\|,
            & {\bf t}({\bf b})\neq{\bf 0},\\
        {\bf 0},
            & {\bf t}({\bf b})={\bf 0},
    \end{cases}
\end{equation}
where
\begin{equation}
    {\bf t}({\bf b})
    :=
    \sum_{i\in[L]}(-1)^{b_i}{\bf n}_i,
\end{equation}
and each ${\bf n}_i$ is a unit vector.
Since $\|{\bf n}_i\|=1$, we have
\begin{align}
    P
    &\ge
    \frac{1}{2}
    +
    \frac{1}{2L}
    \sum_{i\in[L]}
    {\bf n}_i\cdot{\boldsymbol\mu}_i \notag\\
    &=
    \frac{1}{2}
    +
    \frac{1}{2L}
    \mathbb{E}_{\bf b}
    \big[
        {\bf t}({\bf b})\cdot{\bf r}({\bf b})
    \big] \notag\\
    &=
    \frac{1}{2}
    +
    \frac{1}{2L}
    \mathbb{E}_{\bf b}
    \big[
        \|{\bf t}({\bf b})\|
    \big].
\end{align}
In this representation, the remaining design parameters are the unit vectors $\{{\bf n}_i\}_{i\in[L]}$.

\paragraph{Classical setting}
Consider the choice
\begin{equation}
    {\bf n}_i={\bf w}_{i\bmod 2},
\end{equation}
where
\begin{equation}
    {\bf w}_0=
    \begin{pmatrix}
        0 & 0 & 1
    \end{pmatrix}^{\mathsf{T}},
    \qquad
    {\bf w}_1=
    \begin{pmatrix}
        0 & 0 & -1
    \end{pmatrix}^{\mathsf{T}}.
\end{equation}
Then
\begin{equation}
    {\bf t}({\bf b})
    =
    \begin{pmatrix}
        0 & 0 & x({\bf b})
    \end{pmatrix}^{\mathsf{T}},
\end{equation}
where
\begin{equation}
    x({\bf b})
    :=\sum_{i:\text{even}}(-1)^{b_i}+\sum_{i:\text{odd}}(-1)^{b_i+1}
    =\sum_{i\in[L]}(-1)^{b_i+i}
\end{equation}
Thus the state is diagonal in the computational basis:
\begin{equation}
    \rho({\bf b})
    =
    \begin{cases}
        \frac{1}{2}(\mathbb{I}+\texttt{Z}),
            & x({\bf b})>0,\\
        \frac{1}{2}\mathbb{I},
            & x({\bf b})=0,\\
        \frac{1}{2}(\mathbb{I}-\texttt{Z}),
            & x({\bf b})<0.
    \end{cases}
\end{equation}
Since multiplication by the deterministic signs $(-1)^i$ does not change the distribution of the Rademacher sum,
\begin{equation}
    \sum_{{\bf b}\in\{0,1\}^L}
    \left|
        \sum_{i\in[L]}(-1)^{b_i+i}
    \right|
    =
    \sum_{{\bf b}\in\{0,1\}^L}
    \left|
        \sum_{i\in[L]}(-1)^{b_i}
    \right|.
\end{equation}
Therefore,
\begin{align}
    \mathbb{E}_{\bf b}
    \big[
        \|{\bf t}({\bf b})\|
    \big]
    &=
    \mathbb{E}_{\bf b}
    \big[
        |x({\bf b})|
    \big] \notag\\
    &=
    \mathbb{E}_{X}
    \big[
        |L-2X|
    \big], \quad X\sim\operatorname{Bin}\left(L,\tfrac{1}{2}\right) \notag\\
    &=
    \frac{1}{2^L}
    \sum_{\ell=0}^{L}
    \binom{L}{\ell}|L-2\ell| \notag\\
    &=
    \frac{L}{2^{L-1}}
    \binom{L-1}{\lfloor L/2\rfloor}
    =:t_{\rm C}.
\end{align}
Consequently, the lower bound of the average decoding success probability under the Helstrom measurements is
\begin{equation}
    P
    \ge
    \frac{1}{2}
    +
    \frac{1}{2^{L}}
    \binom{L-1}{\lfloor L/2\rfloor}.
\end{equation}
This value coincides with the classical upper bound in \eqref{eq:L1rac}.
Hence, this classical construction is optimal.

\paragraph{Quantum setting}
Next, consider the tetrahedral choice
\begin{equation}
    {\bf n}_i={\bf u}_{i\bmod 4},
\end{equation}
where
\begin{align}
    {\bf u}_0
    &=
    \frac{1}{\sqrt{3}}
    \begin{pmatrix}
        \phantom{-}1 & \phantom{-}1 & \phantom{-}1
    \end{pmatrix}^{\mathsf{T}},
    &
    {\bf u}_1
    &=
    \frac{1}{\sqrt{3}}
    \begin{pmatrix}
        \phantom{-}1 & -1 & -1
    \end{pmatrix}^{\mathsf{T}}, \notag\\
    {\bf u}_2
    &=
    \frac{1}{\sqrt{3}}
    \begin{pmatrix}
        -1 & \phantom{-}1 & -1
    \end{pmatrix}^{\mathsf{T}},
    &
    {\bf u}_3
    &=
    \frac{1}{\sqrt{3}}
    \begin{pmatrix}
        -1 & -1 & \phantom{-}1
    \end{pmatrix}^{\mathsf{T}}.
\end{align}
Define
\begin{equation}
    X_g({\bf b})
    :=
    \sum_{i\in[L]:\, i\bmod 4=g}
    (-1)^{b_i},
    \qquad g\in\{0,1,2,3\}.
\end{equation}
Then
\begin{equation}
    {\bf t}({\bf b})
    =
    \frac{1}{\sqrt{3}}
    \begin{pmatrix}
        A({\bf b}) & B({\bf b}) & C({\bf b})
    \end{pmatrix}^{\mathsf{T}},
\end{equation}
where
\begin{align}
    A({\bf b})
    &:=
    X_0({\bf b})+X_1({\bf b})-X_2({\bf b})-X_3({\bf b}), \notag\\
    B({\bf b})
    &:=
    X_0({\bf b})-X_1({\bf b})+X_2({\bf b})-X_3({\bf b}),\\
    C({\bf b})
    &:=
    X_0({\bf b})-X_1({\bf b})-X_2({\bf b})+X_3({\bf b}). \notag
\end{align}
Each of $A({\bf b})$, $B({\bf b})$, and $C({\bf b})$ has the same distribution as $\sum_{i\in[L]}(-1)^{b_i}$. Hence,
\begin{equation}
    \mathbb{E}_{\bf b}\big[|A({\bf b})|\big]
    =
    \mathbb{E}_{\bf b}\big[|B({\bf b})|\big]
    =
    \mathbb{E}_{\bf b}\big[|C({\bf b})|\big]
    =
    t_{\rm C}.
\end{equation}
By the Cauchy--Schwarz inequality,
\begin{align}
    \mathbb{E}_{\bf b}
    \big[
        \|{\bf t}({\bf b})\|
    \big]
    &=
    \mathbb{E}_{\bf b}
    \left[
        \sqrt{
            \frac{
                A({\bf b})^2+B({\bf b})^2+C({\bf b})^2
            }{3}
        }
    \right] \notag\\
    &\ge
    \frac{1}{3}
    \mathbb{E}_{\bf b}
    \big[
        |A({\bf b})|
        +|B({\bf b})|
        +|C({\bf b})|
    \big] \notag\\
    &=
    t_{\rm C}.
\end{align}
Moreover, the inequality is strict for $L>1$.
Indeed, equality would require $|A({\bf b})|=|B({\bf b})|=|C({\bf b})|$ for all ${\bf b}\in\{0,1\}^L$.
However, for ${\bf b}=0\dots011$,
\begin{equation}
    (A({\bf b}),B({\bf b}),C({\bf b}))=\left\{
    \begin{array}{ll}
         (-4,0,0) & L\equiv 0\pmod{4}  \\
         (-3,1,1) & L\equiv 1\pmod{4}  \\
         (-2,0,0) & L\equiv 2\pmod{4}  \\
         (-3,1,-1) & L\equiv 3\pmod{4}  \\
    \end{array}
    \right.
\end{equation}
holds for $L>1$.
Therefore, there exits ${\bf b}\in\{0,1\}^L$ such that $|A({\bf b})|=|B({\bf b})|=|C({\bf b})|$ does not hold.
Hence,
\begin{equation}
    \mathbb{E}_{\bf b}
    \big[
        \|{\bf t}({\bf b})\|
    \big]
    >
    t_{\rm C}.
\end{equation}
Consequently, this quantum construction satisfies
\begin{equation}
    P
    >
    \frac{1}{2}
    +
    \frac{1}{2^{L}}
    \binom{L-1}{\lfloor L/2\rfloor},
\end{equation}
and hence strictly outperforms the optimal classical $(L,1)$-RAC for every $L>1$.

Finally, we describe a concrete method for computing
${\boldsymbol\mu}_i$, which is required for evaluating the average
decoding success probability, together with its computational cost.
We first rewrite ${\bf t}({\bf b})$ by introducing random
variables as follows:
\begin{align}
    {\bf t}({\bf b})
    &=\sum_{g\in\{0,1,2,3\}}\sum_{i\in G_g}(-1)^{b_i}{\bf n}_i\notag\\
    &=\sum_{g\in\{0,1,2,3\}}{\bf u}_gX_g\notag\\
    &=:{\bf t}(Y).
\end{align}
Here,
\begin{equation}
    G_g:=\{i\in[L]:i\pmod{4}=g\}
\end{equation}
and
\begin{equation}
    Y:=(X_0,X_1,X_2,X_3)
\end{equation}
is a four-dimensional random variable.
Since the random variables $X_g$ are independent, we have
\begin{align}
    \Pr(Y=(x_0,x_1,x_2,x_3))
    =
    \prod_{g\in\{0,1,2,3\}}\Pr(X_g=x_g).
\end{align}
Moreover, since $X_g$ is the sum of $m_g:=|G_g|$ independent
$\pm 1$-valued random variables,
\begin{align}
    \Pr(X_g=x_g)
    =
    2^{-m_g}
    \binom{m_g}{\frac{m_g-x_g}{2}}
\end{align}
for $x_g\in\{-m_g,-m_g+2,\ldots,m_g\}$, and it is zero otherwise.
Similarly, we denote by ${\bf r}(Y)$ the expression obtained by applying
the same change of variables to ${\bf r}({\bf b})$.
Now consider an index $i_g\in G_g$.
For bit strings ${\bf b}$ satisfying $Y=(x_0,x_1,x_2,x_3)$, the conditional average of
$(-1)^{b_{i_g}}$ depends, by symmetry, only on $g$.
Hence, for some constant $c_g$, we may write
\begin{equation}
    \sum_{{\bf b}:X_0=x_0,\dots,X_3=x_3}
    \Pr({\bf b}\mid x_0,\dots,x_3)(-1)^{b_{i_g}}
    =
    c_g .
\end{equation}
Summing both sides over $i_g\in G_g$, the left-hand side becomes
\begin{align}
    &\sum_{i_g\in G_g}
    \sum_{{\bf b}:X_0=x_0,\dots,X_3=x_3}
    \Pr({\bf b}\mid x_0,\dots,x_3)(-1)^{b_{i_g}}
    \notag\\
    &=
    \sum_{{\bf b}:X_0=x_0,\dots,X_3=x_3}
    \Pr({\bf b}\mid x_0,\dots,x_3)X_g
    \notag\\
    &=
    x_g .
\end{align}
On the other hand, the right-hand side is
\begin{equation}
    \sum_{i_g\in G_g} c_g
    =
    m_g c_g.
\end{equation}
Therefore,
\begin{equation}
    c_g=\frac{x_g}{m_g}.
\end{equation}
Using this relation, for $i\in G_g$, we obtain
\begin{align}
    {\boldsymbol\mu}_i
    &=
    \sum_{x_0,\dots,x_3}
    \Pr(x_0,\dots,x_3)\notag\\
    &\quad\cdot
    \sum_{{\bf b}:X_0=x_0,\dots,X_3=x_3}
    \Pr({\bf b}\mid x_0,\dots,x_3)
    (-1)^{b_i}{\bf r}({\bf b})
    \notag\\
    &=
    \sum_{x_0,\dots,x_3}
    \Pr(x_0,\dots,x_3)
    {\bf r}(x_0,\dots,x_3)
    \frac{x_g}{m_g}.
\end{align}
Thus, ${\boldsymbol\mu}_i$ is identical for all $i\in G_g$. We denote
this common vector by ${\boldsymbol\mu}_g$.
More explicitly, let
\begin{equation}
    \chi_g:=\{-m_g,-m_g+2,\dots,m_g\}.
\end{equation}
Then
\begin{align}
    {\boldsymbol\mu}_g
    &=
    \sum_{x_0\in\chi_0}
    \sum_{x_1\in\chi_1}
    \sum_{x_2\in\chi_2}
    \sum_{x_3\in\chi_3}
    \Pr\big(Y=(x_0,x_1,x_2,x_3)\big)
    \notag\\
    &\quad\cdot
    {\bf r}(x_0,x_1,x_2,x_3)
    \frac{x_g}{m_g}.
\end{align}
The computation of ${\boldsymbol\mu}_g$ is performed by enumerating all
tuples in $\chi_0\times\chi_1\times\chi_2\times\chi_3$.
Since $|\chi_g|=m_g+1$, the number of summation is $\prod_{g\in\{0,1,2,3\}}(m_g+1)$.
Hence, the computational cost of computing ${\boldsymbol\mu}_g$ is
\begin{equation}
    O\!\left(
        \prod_{g\in\{0,1,2,3\}}(m_g+1)
    \right)
    =
    O(L^4).
\end{equation}
\end{proofE}

\begin{figure*}[t]
  \centering
    \includegraphics[scale=1]{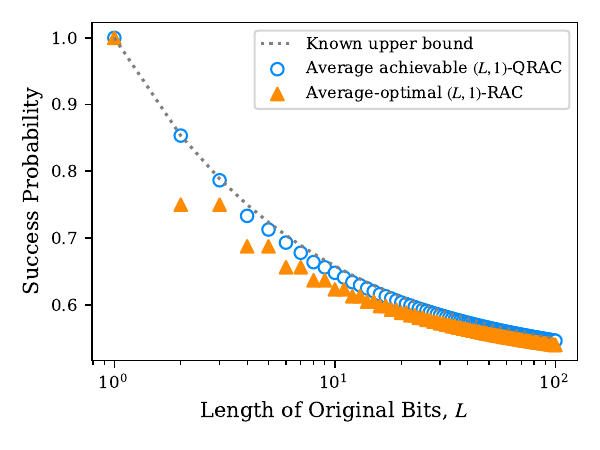}
  \caption{Analytically obtained decoding success probability of $(L,1)$-(Q)RACs, together with the known upper bound~\cite{manvcinska2022geometry}.}
  \label{fig:L1}
\end{figure*}

Figure~\ref{fig:L1} shows the decoding success probabilities of average-optimal $(L,1)$-RACs and the average achievable $(L,1)$-QRACs proposed in this paper.
The figure clearly illustrates the classical--quantum gap, which was already suggested in \cite{ambainis2024quantum}.
However, unlike the previous construction, the proposed quantum construction provides a provable classical--quantum gap for all $L>1$.

The final example of an average-optimal RAC is the case $k=L-1$, which is described below.

\begin{corollaryE}[][normal]
\label{corollary:average}
The average decoding success probability of an $(L,L-1)$-RAC is upper bounded by
\begin{equation}
    \label{eq:LL1-avg}
    1-\frac{1}{2L}.
\end{equation}
\end{corollaryE}
\begin{proofE}[text proof={Proof of Corollary~\ref{corollary:average}}]
Substituting $k=L-1$ into \eqref{eq:average} proves the claim.
\end{proofE}

\begin{corollaryE}[][normal]
\label{corollary:single-parity-check-avg}
The upper bound of Eq.~\eqref{eq:LL1-avg} is achieved by taking $S$ as single parity-check code.
\end{corollaryE}
\begin{proofE}
Since the single parity-check code consists of all even-parity codewords, for any ${\bf b}\in\{0,1\}^L$,
\begin{equation}
    \min_{{\bf s}\in S} d_{\rm H}({\bf b},{\bf s})
    =
    \begin{cases}
        0, & {\bf b}\in S, \\
        1, & {\bf b}\notin S.
    \end{cases}
\end{equation}
Because there are $2^{L-1}$ even-parity points and $2^{L-1}$ odd-parity points,
\begin{equation}
    \sum_{{\bf b}\in\{0,1\}^L}
    \min_{{\bf s}\in S} d_{\rm H}({\bf b},{\bf s})
    =
    2^{L-1}.
\end{equation}
Therefore,
\begin{equation}
    \frac{1}{2^L}
    \sum_{{\bf b}\in\{0,1\}^L}
    \min_{{\bf s}\in S} d_{\rm H/L}({\bf b},{\bf s})
    =
    \frac{1}{2L}.
\end{equation}
Hence, if $S$ is chosen to be the single parity-check code, the corresponding average decoding success probability is
\begin{equation}
    1-\frac{1}{2L},
\end{equation}
which attains the upper bound in~\eqref{eq:LL1-avg}.
\end{proofE}

\subsection{Worst-Case Optimal \texorpdfstring{$(L,k)$}{(L,k)}-RAC}
\label{sec:rac-worst-upper}

Next, we formulate the problem of constructing worst-case optimal $(L,k)$-RAC as follows:

\begin{theoremE}[][normal]
\label{theorem:worst}
Let $k,L\in\mathbb{N}$ satisfy $1\le k\le L$.
The maximum achievable worst-case decoding success probability of an $(L,k)$-RAC is
\begin{equation}
\label{eq:worst-prob}
    1- \min_{\substack{S\subset[0,1]^L\\|S|=2^k}} d_{\mathrm{Haus}}^{\rightarrow}\big(\V,{\rm conv}(S);\,d_\infty\big)
\end{equation}
where
\begin{equation}
    d_{\mathrm{Haus}}^{\rightarrow}(A,B;d):=\sup_{{\bf a}\in A}\operatorname*{inf\vphantom{p}}_{{\bf b}\in B}d({\bf a},{\bf b})
\end{equation}
is a directed Hausdorff distance~\cite{hausdorff1914grundzüge}.
\end{theoremE}
\begin{proofE}
We write the worst-case decoding success probability in terms of the corresponding worst-case error probability $\epsilon_{\rm worst}(P_{\rm E},P_{\rm D})$:
\begin{equation}
\min_{i\in[L]}\min_{{\bf b}\in \V}\Pr(b_i\mid {\bf b})
=1-\epsilon_{\rm worst}(P_{\rm E},P_{\rm D}),
\end{equation}
where
\begin{equation}
\epsilon_{\rm worst}(P_{\rm E},P_{\rm D})
=\max_{i\in[L]}\max_{{\bf b}\in \V}\Pr(B'_i\neq b_i\mid B={\bf b}).
\end{equation}
We aim to lower bound $\min_{P_{\rm E},P_{\rm D}}\epsilon_{\rm worst}(P_{\rm E},P_{\rm D})$.

For a given encoder--decoder pair $(P_{\rm E},P_{\rm D})$, define the expected decoded bit value by
\begin{equation}
    \overline{c}_i({\bf b},P_{\rm E},P_{\rm D})
    =\sum_{b'_i\in\{0,1\}} b'_i\;\Pr(b'_i\mid {\bf b})\in[0,1].
\end{equation}
Since $b_i\in\{0,1\}$, and $\overline{c}_i({\bf b},P_{\rm E},P_{\rm D})=\Pr(B'_i=1\mid B={\bf b})$, we have
\begin{equation}
    \Pr(B'_i\neq b_i\mid B={\bf b})
    =\big|b_i-\overline{c}_i({\bf b},P_{\rm E},P_{\rm D})\big|.
\end{equation}
Therefore,
\begin{align}
\epsilon_{\rm worst}(P_{\rm E},P_{\rm D})
&=\max_{i\in[L]}\max_{{\bf b}\in \V}\big|b_i-\overline{c}_i({\bf b},P_{\rm E},P_{\rm D})\big|\notag\\
&=\max_{{\bf b}\in \V}d_\infty\big({\bf b},\overline{{\bf c}}({\bf b},P_{\rm E},P_{\rm D})\big),\label{eq:e-worst}
\end{align}
where $\overline{{\bf c}}({\bf b},P_{\rm E},P_{\rm D})=(\overline{c}_0,\dots,\overline{c}_{L-1})\in[0,1]^L$.

We now rewrite the minimization over $(P_{\rm E}, P_{\rm D})$ as the problem of selecting a set $S \subset [0,1]^L$ and a mapping $f:\{0,1\}^L \to {\rm conv}(S)$.
To this end, we show the following two statements:
\begin{itemize}
\item For any $(P_{\rm E},P_{\rm D})$, there exist a set $S\subset[0,1]^L$ with $|S|\le 2^k$ and a map
$f:\V\rightarrow {\rm conv}(S)$ such that $\overline{{\bf c}}({\bf b},P_{\rm E},P_{\rm D})=f({\bf b})$ for all ${\bf b}\in \V$.
\item Conversely, for any $S\subset[0,1]^L$ with $|S|\le 2^k$ and any map $f:\V\rightarrow {\rm conv}(S)$, there exist $(P_{\rm E},P_{\rm D})$ such that
$f({\bf b})=\overline{{\bf c}}({\bf b},P_{\rm E},P_{\rm D})$ for all ${\bf b}\in \V$.
\end{itemize}

First, fix a decoder $P_{\rm D}$ and define, for each message ${\bf m}\in\{0,1\}^k$,
\begin{equation}
\label{eq:def-cm}
{\bf c}^{(m)}(P_{\rm D})
:=\sum_{{\bf b}' \in \V} {\bf b}'\, P_{\rm D}({\bf b}'\mid {\bf m})
\in[0,1]^L.
\end{equation}
Collect these points into
\begin{equation}
\label{eq:S-3}
S(P_{\rm D}):=\big\{{\bf c}^{(m)}(P_{\rm D})\bigm| {\bf m}\in \{0,1\}^k\big\},
\end{equation}
which satisfies $|S(P_{\rm D})|\le 2^k$.
Using this notation, the expected decoded string can be rewritten as
\begin{align}
\overline{{\bf c}}({\bf b},P_{\rm E},P_{\rm D})
&=\sum_{{\bf b}'\in \V}{\bf b}'\sum_{{\bf m}\in\{0,1\}^k}P_{\rm D}({\bf b}'\mid {\bf m})P_{\rm E}({\bf m}\mid {\bf b})\notag\\
&=\sum_{{\bf m}\in\{0,1\}^k}{\bf c}^{(m)}(P_{\rm D})\,P_{\rm E}({\bf m}\mid {\bf b})\notag\\
&=: f({\bf b}).
\end{align}
Since the coefficients $P_{\rm E}({\bf m}\mid {\bf b})$ form a probability distribution, $f({\bf b})$ is a convex combination of points in $S(P_{\rm D})$, and hence $f({\bf b})\in {\rm conv}(S(P_{\rm D}))$. This proves the first statement.

For the converse direction, suppose $S=\{{\bf c}^{(1)},{\bf c}^{(2)},\dots\}\subset[0,1]^L$ with $|S|\le 2^k$, and let $f:\V\rightarrow {\rm conv}(S)$.
For each message ${\bf m}$, associate it with a point ${\bf c}^{(m)}\in S$, and define the decoder componentwise by
\begin{equation}
\label{eq:pd-3}
P_{\rm D}(b'_i\mid {\bf m})
=(c^{(m)}_{i})^{b'_i}\,(1-c^{(m)}_{i})^{1-b'_i}.
\end{equation}
Then $\sum_{b'_i\in\{0,1\}} b'_i\,P_{\rm D}(b'_i\mid {\bf m})=c^{(m)}_i$, and hence $\sum_{{\bf b}'\in \V}{\bf b}'\,P_{\rm D}({\bf b}'\mid {\bf m})={\bf c}^{(m)}$.
Since $f({\bf b})\in{\rm conv}(S)$, there exist coefficients $\lambda_{\bf m}({\bf b})\ge 0$ with $\sum_{{\bf m}\in\{0,1\}^k}\lambda_{\bf m}({\bf b})=1$ such that
\begin{equation}
f({\bf b})=\sum_{{\bf m}\in\{0,1\}^k}\lambda_{\bf m}({\bf b})\,{\bf c}^{(m)}.
\end{equation}
Define the encoder by
\begin{equation}
\label{eq:pe-3}
P_{\rm E}({\bf m}\mid {\bf b})=\lambda_{\bf m}({\bf b}).
\end{equation}
Then the resulting expectation satisfies $\overline{{\bf c}}({\bf b},P_{\rm E},P_{\rm D})=f({\bf b})$, proving the second statement.

Combining the above, we obtain
\begin{align}
    &\min_{P_{\rm E},P_{\rm D}} \max_{{\bf b}\in \V}\big\|{\bf b}-\overline{{\bf c}}({\bf b},P_{\rm E},P_{\rm D})\big\|_\infty\notag\\
    &=\min_{\substack{S\subset [0,1]^L\\|S|=2^k}} \max_{{\bf b}\in \V}\min_{{\bf x}\in{\rm conv}(S)}
d_\infty({\bf b},{\bf x})\notag\\
    &=\min_{\substack{S\subset [0,1]^L\\|S|=2^k}} d_{\mathrm{Haus}}^{\rightarrow}\big(\V,{\rm conv}(S);\,d_\infty\big)
\end{align}
where
\begin{equation}
    d_{\mathrm{Haus}}^{\rightarrow}(A,B;d):=\sup_{{\bf a}\in A}\operatorname*{inf\vphantom{p}}_{{\bf b}\in B}d({\bf a},{\bf b})
\end{equation}
is a directed Hausdorff distance.
Thus the maximum worst-case decoding success probability of an $(L,k)$-RAC is
\begin{align}
    &1-\min_{P_{\rm E},P_{\rm D}}\epsilon_{\rm worst}(P_{\rm E},P_{\rm D})\notag\\
    &=1-\min_{\substack{S\subset [0,1]^L\\|S|= 2^k}} d_{\mathrm{Haus}}^{\rightarrow}\big(\V,{\rm conv}(S);\,d_\infty\big).
\end{align}
\end{proofE}
\begin{lemmaE}[][normal]
\label{lemma:worst-rewrite}
The problem
\begin{equation}
    \min_{\substack{S\subset [0,1]^L\\|S|= 2^k}} d_{\mathrm{Haus}}^{\rightarrow}\big(\V,{\rm conv}(S);\,d_\infty\big)    
\end{equation}
can be rewritten as
\begin{equation}
    \max_{\substack{\alpha\in[0,1]\\S\subset [0,1]^L\\|S|= 2^k}} \alpha
    \quad \text{s.t.} \quad
    \alpha\,C \subset P(S) \subset C
\end{equation}
where
\begin{equation}
    C:=[-1,1]^L
\end{equation}
and
\begin{equation}
    P(S):=2\operatorname{conv}(S)-1.
\end{equation}
\end{lemmaE}
\begin{proofE}
We introduce the change of variables
\begin{equation}
    {\bf y}=2{\bf x}-{\bf 1}\in[-1,1]^L,
    \qquad
    {\bf q}=2{\bf b}-{\bf 1}\in\{-1,1\}^L.
\end{equation}
Then, since $|q_i-y_i|=1-q_i y_i$, we can write
\begin{equation}
    d_\infty({\bf b},{\bf x})
    =\max_{i\in[L]}|b_i-x_i|
    =\frac{1-\min_{i\in[L]}q_i y_i}{2}.
\end{equation}
\\
Then,
\begin{align}
    \label{eq:dhaus-rewrite}
    &d^\rightarrow_{\rm Haus}\big(\{0,1\}^L,\operatorname{conv}(S);d_\infty\big)\notag\\
    &=\max_{{\bf b}\in\{0,1\}^L}\min_{{\bf x}\in\operatorname{conv}(S)}d_\infty({\bf b},{\bf x})\notag\\
    &=\frac{1}{2}\big(1-r(P)\big),
\end{align}
where
\begin{equation}
    r(P):=\min_{{\bf q}\in\{\pm1\}^L}\max_{{\bf y}\in P}\min_{i\in[L]}q_i y_i.
\end{equation}
\\
Let
\begin{equation}
    \Delta_L:=\left\{{\boldsymbol\lambda}:\lambda_i\ge0,\ \sum_{i\in[L]}\lambda_i=1\right\},
    \qquad
    u_i:=\lambda_i q_i.
\end{equation}
Then $\|{\bf u}\|_1=1$, and we can rewrite $r(P)$ as
\begin{align}
    r(P)
    &=\min_{{\bf q}\in\{\pm1\}^L}\max_{{\bf y}\in P}\min_{{\boldsymbol\lambda}\in\Delta_L}\sum_{i\in[L]}\lambda_i q_i y_i\notag\\
    &=\min_{{\bf q}\in\{\pm1\}^L}\min_{{\boldsymbol\lambda}\in\Delta_L}\max_{{\bf y}\in P}\sum_{i\in[L]}\lambda_i q_i y_i\notag\\
    &=\min_{\|{\bf u}\|_1=1}\max_{{\bf y}\in P}{\bf u}\cdot{\bf y}\notag\\
    &=\min_{\|{\bf u}\|_1=1}h_P({\bf u}),
\end{align}
where we used the minimax theorem
, and $h_P({\bf u})=\max_{{\bf y}\in P}{\bf u}\cdot{\bf y}$ denotes the support function of $P$.

Let $\beta:=\|{\bf u}\|_1$ and $\widetilde{{\bf u}}:={\bf u}/\beta$.
If
\begin{equation}
    \max_{{\bf y}\in P}\widetilde{{\bf u}}\cdot{\bf y}
    \ge\alpha
\end{equation}
for arbitrary $\widetilde{\bf u}$, then
\begin{equation}
    \label{eq:rPalpha}
    r(P)=\min_{\widetilde{\bf u}}\max_{{\bf y}\in P}\widetilde{\bf u}\cdot{\bf y}\ge\alpha
\end{equation}
and
\begin{equation}
    h_P({\bf u})
    =\max_{{\bf y}\in P}\beta\,\widetilde{{\bf u}}\cdot{\bf y}
    \ge \beta\alpha
    =\alpha\|{\bf u}\|_1.
\end{equation}
On the other hand, from the definition of $h_P({\bf u})$,
\begin{align}
    h_{\alpha C}({\bf u})
    &=\max_{{\bf y}\in \alpha[-1,1]^L}{\bf u}\cdot{\bf y}\notag\\
    &=\max_{|y_i|\le\alpha}\sum_{i\in[L]}u_iy_i\notag\\
    &=\sum_{i\in[L]}|u_i|\alpha\notag\\
    &=\alpha\|{\bf u}\|_1.
\end{align}
Hence,
\begin{equation}
    \label{eq:aC}
    \alpha C \subset P.
\end{equation}
From the definition of $P$ and $C$, it is obvious
\begin{equation}
    \label{eq:PC}
    P \subset C.
\end{equation}
Therefore, we obtain
\begin{equation}
    \label{eq:aCPC}
    \alpha\,C \subset P \subset C.
\end{equation}
The original objective is finding $S$ that minimizes $d^\rightarrow_{\rm Haus}\big(\{0,1\}^L,\operatorname{conv}(S);d_\infty\big)=\frac{1}{2}\big(1-r(P)\big)$ (Eq.~\eqref{eq:dhaus-rewrite}) where $r(P)\ge\alpha$ (Eq.~\eqref{eq:rPalpha}).
Hence, the problem can be rewritten as finding maximum $\alpha$ that satisfies Eq.~\eqref{eq:aCPC}.
\end{proofE}

By Lemma~\ref{lemma:worst-rewrite}, the problem of maximizing the worst-case decoding probability of an $(L,k)$-RAC is, up to a constant factor, equivalent to the following: select a set $S$ of $2^k$ points from the vertices, edges, and interior of the origin-centered $L$-dimensional unit hypercube $Q_L$, form the convex hull of $S$, and then maximize the volume of an axis-aligned hypercube, concentric with $Q_L$, that is contained within this convex hull.
This reformulation enables us to establish the following construction of worst-case optimal $(2^k-1,k)$-RACs.

\begin{theoremE}[][normal]
\label{theorem:worst-2k1}
The worst-case decoding success probability of an $(L,k)$-RAC with $L=2^k-1$ is upper bounded by
\begin{equation}
    \label{eq:worst-2k1-upper}
    \frac{1}{2}+\frac{1}{2L}.
\end{equation}
\end{theoremE}
\begin{proofE}
Combining equation~\eqref{eq:aC}, after dividing both sides by $\alpha > 0$, with equation~\eqref{eq:PC}, we obtain
\begin{equation}
    P \subset C \subset \frac{1}{\alpha} P.
\end{equation}
\\
When $2^k=L+1$, the set $S$ consists of $L+1$ points.
If these $L+1$ points are not affinely independent, then the volume of $\operatorname{conv}(S)$ becomes zero; hence, we restrict our attention to the case in which they are affinely independent.
If these $L+1$ points are affinely independent, then $\operatorname{conv}(S)$ is an $L$-simplex, and so is $P$.
In addition, it is obvious that $C$ is centrally-symmetric convex bodies.
By \cite[Corollary~5.8]{gordon2004john}, if $T$ is a non-degenerate $L$-simplex and $C$ is a centrally-symmetric convex body satisfying
\begin{equation}
    T \subset C \subset t\,T,
\end{equation}
then $t\ge L$.
Substituting $T=P$ and $t=1/\alpha$, we obtain $1/\alpha\ge L$, i.e., $r(P)\le 1/L$.
\\
Therefore,
\begin{equation}
    d^\rightarrow_{\rm Haus}\big(\{0,1\}^L,\operatorname{conv}(S);d_\infty\big)
    \ge \frac{1}{2}\left(1-\frac{1}{L}\right),
\end{equation}
and hence, when $L=2^k-1$, the worst-case decoding success probability is at most
\begin{equation}
    \frac{1}{2}+\frac{1}{2L}.
\end{equation}
\end{proofE}
\begin{theoremE}[][normal]
\label{theorem:worst-2k1-b}
The upper bound of Eq.~\eqref{eq:worst-2k1-upper} is achieved by taking $S$ as binary simplex code.
\end{theoremE}
\begin{proofE}
The binary simplex code $S$ can be written as
\begin{equation}
    S
    =
    \left\{
    \bigl(\langle {\bf u},{\bf y}\rangle\bigr)_{{\bf y}\in\Fkx}
    \bigm|
    {\bf u}\in\{0,1\}^k
    \right\},
\end{equation}
\cite{macwilliams1977theory} where the inner products are taken over $\mathbb F_2$.
Let $\mathbf s^{(\mathbf u)}$ denote the codeword of $S$ corresponding to
$\mathbf u$, and write its $\mathbf y$-th coordinate as
\[
    s^{(\mathbf u)}_{\mathbf y}
    =
    \langle \mathbf u,\mathbf y\rangle .
\]
Define
\begin{equation}
    \chi^{(\mathbf u)}_{\mathbf y}
    :=
    1-2s^{(\mathbf u)}_{\mathbf y}
    =
    (-1)^{\langle \mathbf u,\mathbf y\rangle}
    \in\{\pm1\},
\end{equation}
and let
\begin{equation}
    X
    :=
    \left\{
    \chi^{(\mathbf u)}
    :
    \mathbf u\in\{0,1\}^k
    \right\}.
\end{equation}
By construction, if $\mathbf t\in\operatorname{conv}(X)$, then
$(\mathbf 1-\mathbf t)/2\in\operatorname{conv}(S)$.
\\
We first prove the upper bound on
$d^\rightarrow_{\rm Haus}\big(\{0,1\}^L,\operatorname{conv}(S);d_\infty\big)$.
Fix an arbitrary $\mathbf b\in\{0,1\}^L$, and define
\[
    c_{\mathbf y}:=1-2b_{\mathbf y}\in\{\pm1\}
    \qquad
    (\mathbf y\in\Fkx).
\]
For each $\mathbf u\in\{0,1\}^k$, consider
\begin{equation}
    \lambda_{\mathbf u}
    :=
    \frac{1}{L+1}
    \left(
        1
        +
        \frac{1}{L}
        \sum_{\mathbf y\in\Fkx}
        c_{\mathbf y}\chi^{(\mathbf u)}_{\mathbf y}
    \right).
\end{equation}
Since
\[
    \sum_{\mathbf y\in\Fkx}
    c_{\mathbf y}\chi^{(\mathbf u)}_{\mathbf y}
    \ge -L,
\]
we have $\lambda_{\mathbf u}\ge0$ for every $\mathbf u$.
Moreover, by the orthogonality of the Walsh--Hadamard characters,
\[
    \sum_{\mathbf u\in\{0,1\}^k}\lambda_{\mathbf u}=1.
\]
Thus $\{\lambda_{\mathbf u}\}_{\mathbf u\in\{0,1\}^k}$ is a probability
distribution.
\\
Now consider the corresponding convex combination of the points in $X$.
Using again the Walsh--Hadamard orthogonality, we obtain
\begin{equation}
    \operatorname{conv}(X)\ni
    \sum_{\mathbf u\in\{0,1\}^k}
    \lambda_{\mathbf u}\chi^{(\mathbf u)}
    =
    \frac{\mathbf c}{L}
    =:
    \mathbf q .
\end{equation}
Hence
\begin{equation}
    \mathbf x
    :=
    \frac{\mathbf 1-\mathbf q}{2}
    =
    \frac{\mathbf 1}{2}
    -
    \frac{\mathbf c}{2L}
\end{equation}
belongs to $\operatorname{conv}(S)$. Since
$\mathbf b=(\mathbf 1-\mathbf c)/2$, for every $\mathbf y\in\Fkx$ we have
\begin{equation}
    |b_{\mathbf y}-x_{\mathbf y}|
    =
    \frac{1}{2}
    |c_{\mathbf y}-q_{\mathbf y}|
    =
    \frac{1}{2}
    \left|c_{\mathbf y}-\frac{c_{\mathbf y}}{L}\right|
    =
    \frac{1}{2}-\frac{1}{2L}.
\end{equation}
Therefore, for every $\mathbf b\in\{0,1\}^L$, there exists
$\mathbf x\in\operatorname{conv}(S)$ such that
\[
    d_\infty(\mathbf b,\mathbf x)
    =
    \frac{1}{2}-\frac{1}{2L}.
\]
It follows that
\begin{equation}
    \label{eq:haus-upper}
    d^\rightarrow_{\rm Haus}
    \big(\{0,1\}^L,\operatorname{conv}(S);d_\infty\big)
    \le
    \frac{1}{2}-\frac{1}{2L}.
\end{equation}
\\
We next prove the reverse inequality. Since $S$ is a binary simplex code,
every nonzero codeword has Hamming weight
\[
    2^{k-1}=\frac{L+1}{2}.
\]
Consequently, for every $\mathbf x\in\operatorname{conv}(S)$,
\begin{equation}
    \sum_{\mathbf y\in\Fkx} x_{\mathbf y}
    \le
    \frac{L+1}{2}.
\end{equation}
Now take $\mathbf b=\mathbf 1\in\{0,1\}^L$. If
$d_\infty(\mathbf 1,\mathbf x)\le r$ for some
$\mathbf x\in\operatorname{conv}(S)$, then
$x_{\mathbf y}\ge 1-r$ for all $\mathbf y\in\Fkx$. Hence
\[
    L(1-r)
    \le
    \sum_{\mathbf y\in\Fkx} x_{\mathbf y}
    \le
    \frac{L+1}{2}.
\]
This implies
\begin{equation}
    r\ge \frac{1}{2}-\frac{1}{2L}.
\end{equation}
Therefore,
\begin{equation}
    \label{eq:haus-lower}
    d^\rightarrow_{\rm Haus}
    \big(\{0,1\}^L,\operatorname{conv}(S);d_\infty\big)
    \ge
    \frac{1}{2}-\frac{1}{2L}.
\end{equation}
\\
Combining \eqref{eq:haus-upper} and \eqref{eq:haus-lower}, we conclude that,
when $S$ is chosen as the binary simplex code,
\begin{equation}
    d^\rightarrow_{\rm Haus}
    \big(\{0,1\}^L,\operatorname{conv}(S);d_\infty\big)
    =
    \frac{1}{2}-\frac{1}{2L}.
\end{equation}
That is, the worst-case decoding success probability of corresponding $(2^k-1,k)$-RAC is
\begin{equation}
    \frac{1}{2}+\frac{1}{2L}.
\end{equation}
\end{proofE}
\begin{corollaryE}[][normal]
\label{corollary:worst-2k1}
One choice of states and POVMs for the worst-case optimal $(L,k)$-RAC with $L=2^k-1$ is given by Eqs.~\eqref{eq:improved-rho} and \eqref{eq:improved-povm}, respectively.
\end{corollaryE}
\begin{proofE}
The worst-case decoding success probability corresponding to the states and POVMs defined in Eqs.~\eqref{eq:improved-rho} and \eqref{eq:improved-povm} is given by Eq.~\eqref{eq:improved-prob}, and it attains the upper bound established in Theorem~\ref{theorem:worst-2k1} when $L=2^k-1$.
\end{proofE}

Note that \cite{liabotro2017improved} establishes that the worst-case decoding success probability of the optimal \textit{parity-oblivious} $(L,k)$-RAC with $L=2^k-1$ is $\frac{1}{2}+\frac{1}{2L}$, which is the same as Eq.~\eqref{eq:worst-2k1-upper}. 
Our results show that removing the \textit{parity-oblivious} constraint does not increase this upper bound.

Next, we show the existence of a $(2^k-1,k)$-QRAC whose worst-case decoding success probability is strictly higher than that of the worst-case optimal $(2^k-1,k)$-RAC, although its optimality remains unknown.

\begin{theoremE}[][normal]
\label{theorem:worst-2k1-Q}
There exists $(2^k-1,k)$-QRAC whose worst-case decoding success probability is
\begin{equation}
    \frac{1}{2}+\frac{1}{2\sqrt{L}},
\end{equation}
which is strictly greater than that of worst-case optimal $(2^k-1,k)$-RAC for $k>1$.
\end{theoremE}
\begin{proofE}
Let
\begin{equation}
    \rho(\mathbf b)=|\psi(\mathbf b)\rangle\langle\psi(\mathbf b)|,
\end{equation}
where
\begin{equation}
    |\psi(\mathbf b)\rangle=\frac{1}{\sqrt{2}}|0\rangle^{\otimes k}+\frac{1}{\sqrt{2L}}\sum_{i\in[L]}(-1)^{b_i}|i+1\rangle .
\end{equation}
For each $i\in[L]$ and $a\in\{0,1\}$, define
\begin{equation}
    E^a_i=\frac{1}{2}\left(
        \mathbb{I}^{\otimes k}+(-1)^a\bigl(|0\rangle\langle i+1|+|i+1\rangle\langle 0|\bigr)
    \right).
\end{equation}
We first verify that $\rho(\mathbf b)$ is a valid quantum state. Since
$\rho(\mathbf b)$ is a rank-one projector, it is Hermitian and positive
semidefinite. Moreover,
\begin{equation}
    \operatorname{tr}(\rho(\mathbf b))=\langle\psi(\mathbf b)|\psi(\mathbf b)\rangle
    =\frac{1}{2}+\frac{1}{2L}\sum_{i\in[L]}1
    =1.
\end{equation}
Thus $\rho(\mathbf b)$ is a quantum state.
We next verify that $\{E^0_i,E^1_i\}$ is a valid POVM. It is immediate that
\begin{equation}
    (E^a_i)^\dagger = E^a_i
\end{equation}
for each $a\in\{0,1\}$, and
\begin{equation}
    E^0_i+E^1_i=\mathbb{I}^{\otimes k}.
\end{equation}
Let
\begin{equation}
    A_i:=|0\rangle\langle i+1|+|i+1\rangle\langle 0|.
\end{equation}
Since $|i\rangle\neq |0\rangle$, we have
\begin{equation}
    A_i(|0\rangle\pm |i+1\rangle)=\pm (|0\rangle\pm |i+1\rangle).
\end{equation}
Moreover, $A_i$ acts as zero on the orthogonal complement of
$\operatorname{span}\{|0\rangle,|i+1\rangle\}$. Hence the eigenvalues of
$A_i$ are
\begin{equation}
    +1,\ -1,\ 0,\ldots,0.
\end{equation}
Therefore, the eigenvalues of $E^a_i$ are
\begin{equation}
    1,\ 0,\ \frac{1}{2},\ldots,\frac{1}{2},
\end{equation}
and hence $E^a_i$ is positive semidefinite for each $a\in\{0,1\}$.
Thus $\{E^0_i,E^1_i\}$ is a valid two-outcome POVM.
Finally, for the outcome $b_i$, we have
\begin{align}
    \operatorname{tr}\big(E^{b_i}_{i}\rho(\mathbf b)\big)
    &=\langle\psi(\mathbf b)|E^{b_i}_{i}|\psi(\mathbf b)\rangle \notag\\
    &=\frac{1}{2}+\frac{1}{2}(-1)^{b_i}\langle\psi(\mathbf b)|A_i|\psi(\mathbf b)\rangle .
\end{align}
Since
\begin{equation}
    \langle\psi(\mathbf b)|A_i|\psi(\mathbf b)\rangle=\frac{(-1)^{b_i}}{\sqrt{L}},
\end{equation}
we obtain
\begin{equation}
    \operatorname{tr}\big(E^{b_i}_{i}\rho(\mathbf b)\big)=\frac{1}{2}+\frac{1}{2\sqrt{L}}.
\end{equation}
\end{proofE}

\begin{figure*}[t]
  \centering
    \includegraphics[scale=1]{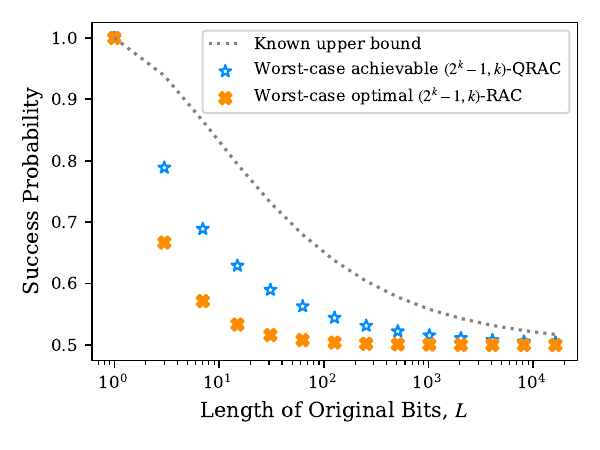}
  \caption{Analytically obtained decoding success probability of $(2^k-1,k)$-(Q)RACs,  together with the known upper bound~\cite{ambainis1999dense,nayak1999optimal}.}
  \label{fig:worst-exact-bound}
\end{figure*}

Theorem~\ref{theorem:worst-2k1-Q} shows the existence of a classical--quantum gap for $(L,k)=(2^k-1,k)$.
Figure~\ref{fig:worst-exact-bound} compares the worst-case optimal $(2^k-1,k)$-RAC, an achievable $(2^k-1,k)$-QRAC, and the known upper bounds on their worst-case decoding success probabilities~\cite{ambainis1999dense,nayak1999optimal}.
Although the QRAC construction is not optimal, the figure clearly demonstrates a classical--quantum gap for this family of parameters.

To further analyze worst-case optimality for other families of parameters, we introduce the following conjecture.

\begin{conjecture}
\label{conjecture:worst}
An optimal set $S$ of the problem Eq.~\eqref{eq:worst-prob} can be chosen so that $S\subset\{0,1\}^L$.
\end{conjecture}
The reduced problem can be written as follows:
\begin{equation}
\label{eq:min-hausdorff-reduced}
    \min_{\substack{S\subset\{0,1\}^L\\|S|=2^k}} d_{\mathrm{Haus}}^{\rightarrow}\big(\V,{\rm conv}(S);\,d_\infty\big).
\end{equation}

\begin{propositionE}[][normal]
\label{conjecture:worst-2}
The following statements are equivalent.
\begin{enumerate}[label=(\roman*)]
  \item Conjecture~\ref{conjecture:worst}.
  \item \label{item:reform}
  An optimal decoder for an $(L,k)$-RAC that maximizes the worst-case decoding success probability can be chosen to be deterministic.
\end{enumerate}
\end{propositionE}
\begin{proofE}
Imposing $S\subset\{0,1\}^L$ is equivalent to requiring ${\bf c}^{(m)}\in\{0,1\}^L$ for all messages ${\bf m}$ (see Eq.~\eqref{eq:S-3}).
Moreover, when ${\bf c}^{(m)}\in\{0,1\}^L$, Eq.~\eqref{eq:pd-3} reduces to
\begin{equation}
P_{\rm D}({\bf b}'\mid {\bf m})=
\left\{
\begin{array}{ll}
1 & {\bf b}'={\bf c}^{(m)},\\
0 & \text{otherwise},
\end{array}
\right.
\end{equation}
i.e., the decoder becomes deterministic.
Conversely, for any deterministic decoder, ${\bf c}^{(m)}(P_{\rm D})$ necessarily lies in $\{0,1\}^L$ (see Eq.~\eqref{eq:def-cm}).
Hence, the set $S(P_{\rm D})=\{{\bf c}^{(m)}(P_{\rm D})\mid {\bf m}\in\{0,1\}^k\}$ is a subset of $\{0,1\}^L$.
Therefore, assuming $S\subset\{0,1\}^L$ is equivalent to assuming a deterministic decoder.
\end{proofE}

If Conjecture~\ref{conjecture:worst} holds, the problem of finding the optimal $(L,k)$-RAC maximizing the worst-case decoding success probability can be written in MILP.
If not, the solution of such an MILP gives an achievable $(L,k)$-RAC maximizing the worst-case decoding success probability because the feasible region with $S\subset \{0,1\}^L$ is narrower than that with $S\subset[0,1]^L$.
In addition, assuming this conjecture holds, a worst-case optimal $(L,L-1)$-RAC can be constructed as follows.

\begin{corollaryE}[][normal]
\label{theorem:worst2}
Assuming Conjecture~\ref{conjecture:worst} holds, the worst-case decoding success probability of an $(L,L-1)$-RAC is upper bounded by
\begin{equation}
    \label{eq:LL1-rac-upper}
    1-\frac{1}{L}.
\end{equation}
\end{corollaryE}
\begin{proofE}
It suffices to show that, for any $S\subset\{0,1\}^L$,
\begin{equation}
    \label{eq:LL1-lower}
    \max_{{\bf b}\in\{0,1\}^L}
    \min_{{\bf x}\in\operatorname{conv}(S)}
    d_\infty({\bf b},{\bf x})
    \ge \frac{1}{L}.
\end{equation}
Take an arbitrary ${\bf b}\notin S$ and ${\bf x}\in\operatorname{conv}(S)$. Since ${\bf x}\in\operatorname{conv}(S)$, there exist coefficients $\{\lambda_{\bf s}\}_{{\bf s}\in S}$ satisfying $\lambda_{\bf s}\ge0$ and $\sum_{{\bf s}\in S}\lambda_{\bf s}=1$ such that
\begin{equation}
    {\bf x}=\sum_{{\bf s}\in S}\lambda_{\bf s}{\bf s}.
\end{equation}
Because ${\bf b},{\bf s}\in\{0,1\}^L$, we have
\begin{equation}
    \|{\bf b}-{\bf x}\|_1
    =
    \sum_{{\bf s}\in S}\lambda_{\bf s}d_{\rm H}({\bf b},{\bf s}).
\end{equation}
Moreover, since ${\bf b}\notin S$, we have $d_{\rm H}({\bf b},{\bf s})\ge1$ for all ${\bf s}\in S$. Therefore,
\begin{equation}
    \|{\bf b}-{\bf x}\|_1 \ge 1.
\end{equation}
Using $\|{\bf y}\|_\infty \ge \|{\bf y}\|_1/L$, we obtain
\begin{equation}
    d_\infty({\bf b},{\bf x})
    \ge
    \frac{1}{L}\|{\bf b}-{\bf x}\|_1
    \ge
    \frac{1}{L}.
\end{equation}
Hence,
\begin{equation}
    \max_{{\bf b}\in\{0,1\}^L}
    \min_{{\bf x}\in\operatorname{conv}(S)}
    d_\infty({\bf b},{\bf x})
    \ge
    \frac{1}{L}.
\end{equation}
\end{proofE}

\begin{corollaryE}[][normal]
\label{corollary:single-parity-check}
The conjectured upper bound of Eq.~\eqref{eq:LL1-rac-upper} is achieved by taking $S$ as single parity-check code.
\end{corollaryE}
\begin{proofE}
The single parity-check code is given by~\cite{macwilliams1977theory}
\begin{equation}
    \label{eq:single-parity-check-code}
    S=
    \Big\{
        {\bf b}\in \V
        \ \Big|\
        \sum_{i\in[L]} b_i \equiv 0 \pmod{2}
    \Big\}.
\end{equation}
If ${\bf b}\in S$, then ${\bf b}\in\operatorname{conv}(S)$, and hence there exists ${\bf x}\in\operatorname{conv}(S)$ such that $d_\infty({\bf b},{\bf x})=0$.
That is, for every ${\bf b}\in S$,
\begin{equation}
    \min_{{\bf x}\in\operatorname{conv}(S)}
    d_\infty({\bf b},{\bf x})
    =
    0.
\end{equation}
Thus, it remains to consider the case ${\bf b}\notin S$.
For ${\bf b}\notin S$, each vector ${\bf b}\oplus{\bf e}_i$ belongs to $S$, where ${\bf e}_i$ denotes the vector whose $i$-th coordinate is $1$ and whose other coordinates are $0$. Define
\begin{equation}
    {\bf x}_{\bf b}
    :=
    \frac{1}{L}
    \sum_{i\in[L]}({\bf b}\oplus{\bf e}_i)
    \in \operatorname{conv}(S).
\end{equation}
Then, for every coordinate $i\in[L]$,
\begin{equation}
    |({\bf x}_{\bf b}-{\bf b})_i|
    =
    \frac{1}{L}.
\end{equation}
Therefore,
\begin{equation}
    d_\infty({\bf b},{\bf x}_{\bf b})
    =
    \frac{1}{L}.
\end{equation}
Hence, for every ${\bf b}\notin S$,
\begin{equation}
    \min_{{\bf x}\in\operatorname{conv}(S)}
    d_\infty({\bf b},{\bf x})
    \le
    d_\infty({\bf b},{\bf x}_{\bf b})
    =
    \frac{1}{L}.
\end{equation}
That is,
\begin{equation}
    \max_{{\bf b}\in\{0,1\}^L}\min_{{\bf x}\in\operatorname{conv}(S)}
    d_\infty({\bf b},{\bf x})
    \le
    \frac{1}{L}.
\end{equation}
Together with the lower bound in~\eqref{eq:LL1-lower}, this shows that the single parity-check code is optimal under Conjecture~\ref{conjecture:worst}.
\end{proofE}

\subsubsection{Constructing worst-case achievable \texorpdfstring{$(L,k)$}{(L,k)}-RAC}
\label{sec:rac-worst-construct}
To obtain a worst-case optimal $(L,k)$-RAC, one needs to solve a bilinear programming problem \eqref{eq:worst-prob}.
Nevertheless, when Conjecture~\ref{conjecture:worst} is accepted, an achievable solution can be obtained by solving an MILP.
Specifically, consider
\begin{align}
    & \text{minimize} && h \notag\\
    & \text{subject to} && \sum_{{\bf a}\in \V}z_{\bf a}=2^k,\notag\\
    & \phantom{\text{subject to}} && \sum_{{\bf a}\in \V}\lambda_{\bf a}({\bf b})=1 && \hspace{-20mm}\forall {\bf b}\in \V,\notag\\
    & \phantom{\text{subject to}} &&
    -h\le b_i-\sum_{{\bf a}\in \V}\lambda_{\bf a}({\bf b})a_i \le h
    && \notag\\
    & && && \hspace{-20mm}\forall {\bf b}\in \V,\;\forall i\in [L],\notag\\
    & \phantom{\text{subject to}} && z_{\bf a}\in\{0,1\} && \hspace{-20mm}\forall {\bf a}\in \V,\notag\\
    & \phantom{\text{subject to}} && 0\le\lambda_{\bf a}({\bf b})\le z_{\bf a} && \hspace{-20mm}\forall {\bf a}\in \V,\;\forall {\bf b}\in \V.\label{eq:milp-worst}
\end{align}
After solving \eqref{eq:milp-worst}, one can define an encoder and a decoder by setting
\begin{equation}
    P_{\rm E}({\bf m}^{(j)}\mid {\bf b})=\lambda_{{\bf s}^{(j)}}({\bf b}),
\end{equation}
and
\begin{equation}
    P_{\rm D}({\bf b}'\mid {\bf m}^{(j)})=\left\{
    \begin{array}{ll}
        1 & {\bf b}'={\bf s}^{(j)},\\
        0 & \text{otherwise}.
    \end{array}
    \right.
\end{equation}

One may also construct a worst-case achievable $(L,k)$-RAC directly from a set $S$.
Although finding an $S$ that maximizes the worst-case success probability takes an exponentially large time in general, if an optimal $S$ is available, then one can proceed as follows.
Let ${\bf b}^{*(j)}$ denote the $j$-th element of $S$, and let ${\bf m}^{(j)}$ denote the $j$-th element of $\{0,1\}^k$.
Define the deterministic decoder by
\begin{equation}
    P_{\rm D}({\bf b}^{*(j)}\mid {\bf m}^{(j)})=\left\{
    \begin{array}{ll}
        1 & D({\bf m}^{(j)})={\bf b}^{*(j)},\\
        0 & \text{otherwise},
    \end{array}
    \right.
\end{equation}
where $D:\{0,1\}^k\rightarrow \{0,1\}^L$ is a map that assigns ${\bf b}^{*(j)}$ to ${\bf m}^{(j)}$.
That is, the $j$-th message ${\bf m}^{(j)}$ is decoded to ${\bf b}^{*(j)}$.
For each input ${\bf b}\in \V$, define the encoder by
\begin{equation}
    P_{\rm E}({\bf m}^{(j)}\mid {\bf b})=\lambda^*_j({\bf b}),
\end{equation}
where ${\boldsymbol\lambda}^*({\bf b})\in\mathbb{R}^{2^k}_+$ is an optimal solution to
\begin{align}
    & \min_{{\boldsymbol\lambda}} \ \max_{i\in[L]}\Big|\sum_{j\in[2^k]}\lambda_j b^{*(j)}_i-b_i\Big| \notag\\
    & \text{subject to} \quad \sum_{j\in[2^k]}\lambda_{j}=1,\qquad
    \lambda_{j}\ge0 \ \ (j\in[2^k]).\label{eq:lambda}
\end{align}
This can be converted into the following linear program:
\begin{align}
    & \min_{t,\{\lambda_{j}\}} &&t \notag\\
    & \text{subject to} && \sum_{j\in[2^k]}\lambda_{j}b^{*(j)}_i-b_i\le t && \forall i\in[L],\notag\\
    & \phantom{\text{subject to}} && b_i-\sum_{j\in[2^k]}\lambda_{j}b^{*(j)}_i\le t && \forall i\in[L],\notag\\
    & \phantom{\text{subject to}} && \sum_{j\in[2^k]}\lambda_{j}=1,\notag\\
    & \phantom{\text{subject to}} && \lambda_{j}\ge0 && \forall j\in[2^k].
\end{align}
Since this is a linear program, ${\boldsymbol\lambda}^*({\bf b})$ can be computed efficiently.


Here, it is noteworthy that both the average-optimal $(L,L-1)$-RAC and the conjectured worst-case optimal $(L,L-1)$-RAC are realized by the single parity-check code (Corollary~\ref{corollary:single-parity-check-avg} and Corollary~\ref{corollary:single-parity-check}).
In fact, the following construction gives an $(L,L-1)$-RAC that is average-optimal and, assuming Conjecture~\ref{conjecture:worst}, also worst-case optimal.

\begin{theoremE}[][normal]
\label{theorem:upper3}
There exists an $(L,L-1)$-RAC with average decoding success probability $1-\frac{1}{2L}$ and worst-case decoding success probability $1-\frac{1}{L}$, which attain the (conjectured) upper bound.
\end{theoremE}
\begin{proofE}
Let
\begin{equation}
    p({\bf b}):=\bigoplus_{i}b_i
\end{equation}
be a parity of bit string ${\bf b}$.
The claimed performance can be achieved by the state in \eqref{eq:rho-rac} and the POVMs in \eqref{eq:povm-rac} below:
\begin{equation}
\label{eq:rho-rac}
\operatorname{diag}(\rho_{\rm C}({\bf b}))
=
\left\{
\begin{array}{ll}
\displaystyle{|{\bf c}({\bf b})\rangle  }& \displaystyle{\text{for}\quad p({\bf b})=0},\\
\displaystyle{U_{\rm C}|{\bf c}({\bf b})}\rangle& \displaystyle{\text{for}\quad p({\bf b})=1},
\end{array}
\right.
\end{equation}
where
\begin{equation}
\label{eq:}
U_{\rm C}:=\frac{1}{L}\Big(\mathbb{I}^{\otimes L-1}+\sum_{j\in[L-1]}\texttt{X}_j\Big)
\end{equation}
and
\begin{equation}
    {\bf c}({\bf b}):=b_{L-2}\,b_{L-3}\dots b_0
\end{equation}
is the lower $L-1$ bits of ${\bf b}$, and $|x\rangle$ denotes the computational basis vector with a $1$ in the $x$-th position and $0$ elsewhere.
The POVM elements are given by
\begin{equation}
\label{eq:povm-rac}
E^0_i=\frac{1}{2}\Big(\mathbb{I}^{\otimes L-1}+\zeta_i\Big),
\quad
E^1_i=\frac{1}{2}\Big(\mathbb{I}^{\otimes L-1}-\zeta_i\Big),
\end{equation}
with
\begin{equation}
\zeta_i=\left\{
\begin{array}{ll}
    \texttt{Z}_i & \text{for}\quad i<L-1,\\
    \texttt{Z}^{\otimes L-1} &\text{for}\quad  i=L-1.
\end{array}
\right.
\end{equation}
Expressed directly in terms of classical conditional distributions, the above encoder--decoder can be written as follows:
\begin{equation}
P_{\rm E}({\bf m}\mid {\bf b})
= \left\{
\begin{array}{ll}
    1 & \text{for}\quad p({\bf b})=0\ \ \text{and}\ \ {\bf m}={\bf c}({\bf b}),\\[2mm]
    \frac{1}{L} & \text{for}\quad p({\bf b})=1\ \ \text{and}\ \ d_{\rm H}({\bf m},{\bf c}({\bf b}))\le 1,\\[2mm]
    0 & \text{otherwise},
\end{array}
\right.
\end{equation}
and
\begin{equation}
P_{\rm D}(b'_i\mid {\bf m})
= \left\{
\begin{array}{ll}
    1 & \text{for}\quad b'_i=\left\{\begin{array}{ll}m_i & i< L-1\\p({\bf m}) & i=L-1\end{array}\right.,\\[5mm]
    0 & \text{otherwise}.
\end{array}
\right.
\end{equation}

This protocol can be described as follows.
Given ${\bf b}$, if $p({\bf b}) = 1$, flip exactly one bit of ${\bf b}$ uniformly at random; otherwise, do not flip any bit. Let ${\bf m}$ be the resulting lower $k = L - 1$ bits.
For decoding, output $b'_i=m_i$ for $i<L-1$, and output the parity $b'_{L-1}=p({\bf m})$ to estimate $b_{L-1}$.
Under this protocol, for each $i$ the event $b'_i\neq b_i$ occurs with probability $\frac{1}{L}$, while all other positions $j\neq i$ are decoded correctly with probability $1$.
Hence the average decoding success probability equals $1-\big(\frac{1}{2}\cdot0+\frac{1}{2}\cdot\frac{1}{L}\big)=1-\frac{1}{2L}$, and the worst-case decoding success probability equals $1-\frac{1}{L}$, attaining both upper bounds if Conjecture~\ref{conjecture:worst} holds.
\end{proofE}

Furthermore, as shown below, $(L,L-1)$-QRAC attaining the conjectured upper bound can be constructed in a form closely analogous to the encoder and decoder of this $(L,L-1)$-RAC.
Note that the following constructions of the states and POVMs differ from those in~\cite{suzuki2026analytical}, although they achieve the same decoding success probability.

\begin{theoremE}[][normal]
\label{theorem:optimal-qrac}
There exists an $(L,L-1)$-QRAC whose average and worst-case decoding success probabilities are both
\begin{equation}
    \label{eq:LL-1qrac}
    \frac{1}{2}+\frac{1}{2}\sqrt{\frac{L-1}{L}},
\end{equation}
which attains the value of the conjectured bound (Eq.~\eqref{eq:qrac-conjecture}).
\end{theoremE}
\begin{proofE}
The encoding state is given by
\[
\rho_{\rm Q}({\bf b}) = |\psi({\bf b})\rangle\langle\psi({\bf b})|,
\]
where
\begin{equation}
\label{eq:rho-qrac}
|\psi({\bf b})\rangle
=
\begin{cases}
|{\bf c}({\bf b})\rangle, & p({\bf b})=0,\\[1ex]
U_{\rm Q}\,|{\bf c}({\bf b})\rangle, & p({\bf b})=1,
\end{cases}
\end{equation}
and
\begin{equation}
U_{\rm Q}:=\frac{1}{\sqrt{L}}\left(\mathbb{I}^{\otimes k}
+\sum_{j\in[k]}\texttt{X}_j \texttt{Z}_0 \cdots \texttt{Z}_j
\right),
\end{equation}
where $k=L-1$.

Here, we verify that $U_{\rm Q}$ is unitary.
Define
\begin{equation}
    A_j:=\texttt{X}_j \texttt{Z}_0 \cdots \texttt{Z}_{j}.
\end{equation}
Then
\begin{equation}
    A_j^\dagger=-A_j,
    \quad
    A_j^2=-\mathbb{I}^{\otimes k},
    \quad
    A_jA_l=-A_lA_j
    \ \ (j\neq l).
\end{equation}
Define
\begin{equation}
    B := \sum_{j\in[k]}A_j.
\end{equation}
Then
\begin{equation}
    B^\dagger=-B
\end{equation}
and
\begin{align}
    B^2
    &=
    \sum_{j\in[k]}A_j^2
    +
    \sum_{j\neq l}A_jA_l \notag\\
    &=
    -k\;\mathbb{I}^{\otimes k}.
\end{align}
Hence,
\begin{align}
    U^\dagger_{\rm Q}U_{\rm Q}
    &=\frac{1}{L}\Big(\mathbb{I}^{\otimes k}+B\Big)^\dagger\Big(\mathbb{I}^{\otimes k}+B\Big)\notag\\
    &=\frac{1}{L}\Big(\mathbb{I}^{\otimes k}-B^2\Big)\notag\\
    &=\mathbb{I}^{\otimes k}.
\end{align}
Thus, $U_{\rm Q}$ is unitary.

This construction coincides with $\operatorname{diag}(\rho_{\rm C}({\bf b}))$ (see Eq.~\eqref{eq:rho-rac}) except when $p({\bf b})=1$, where the states are modified to ensure mutual orthogonality.

The POVMs are given by
\begin{equation}
    E^0_i=\frac{1}{2}\Big(\mathbb{I}^{\otimes k}+\xi_i\Big),
    \qquad
    E^1_i=\frac{1}{2}\Big(\mathbb{I}^{\otimes k}-\xi_i\Big),
\end{equation}
where
\begin{equation}
\xi_i=
\begin{cases}
\displaystyle
\frac{1}{2}\sqrt{\frac{L}{L-1}}\big(\texttt{Z}_i+U_{\rm Q}\texttt{Z}_iU_{\rm Q}^\dagger\big),
& i<L-1,\\[2ex]
\displaystyle
\frac{1}{2}\sqrt{\frac{L}{L-1}}\big(\texttt{Z}^{\otimes k}-U_{\rm Q}\texttt{Z}^{\otimes k}U^\dagger_{\rm Q}\big),
& i=L-1.
\end{cases}
\end{equation}

We now verify that $\{E_i^0,E_i^1\}$ defines a valid POVM.
Since $E_i^0+E_i^1=\mathbb{I}^{\otimes k}$ is obvious, it remains to show that the minimum eigenvalue of $\xi_i$ is at least $-1$, which implies that both $E_i^0$ and $E_i^1$ are positive semidefinite.
We first consider the case $i<L-1$.
Define
\begin{equation}
    C_i := \sum_{j\neq i}A_j.
\end{equation}
Note the following identities:
\begin{equation}
    \{B,\texttt{Z}_i\}
    =\{A_i,\texttt{Z}_i\}+\Big\{\sum_{j\neq i}A_j,\texttt{Z}_i\Big\}
    =2\sum_{j\neq i}A_j\texttt{Z}_i,
\end{equation}
and
\begin{equation}
    C_i^2
    =(B-A_i)^2
    =(1-k)\mathbb{I}^{\otimes k}.
\end{equation}
Here, we consider the following quantity:
\begin{equation}
    U_{\rm Q}\texttt{Z}U^\dagger_{\rm Q}
    =\frac{1}{L}\Big(\texttt{Z}_i+[B,\texttt{Z}_i]-B\texttt{Z}_iB\Big).
\end{equation}
Noting that $[B,\texttt{Z}_i]=[A_i,\texttt{Z}_i]=2A_i\texttt{Z}_i$, we obtain
\begin{equation}
    \{\texttt{Z}_i,[B,\texttt{Z}_i]\}
    =\{\texttt{Z}_i,2A_i\texttt{Z}_i\}
    =0.
\end{equation}
Moreover,
\begin{align}
    B\texttt{Z}_iB
    &=(A_i+C_i)\texttt{Z}_i(A_i+C_i)\notag\\
    &=A_i\texttt{Z}_iA_i+A_i\texttt{Z}_iC_i+C_i\texttt{Z}_iA_i+C_i^2\notag\\
    &=-A^2_i\texttt{Z}_i-2\texttt{Z}_iA_iC_i+\texttt{Z}C^2_i\notag\\
    &=(2-k)\texttt{Z}_i-2\texttt{Z}_iA_iC_i.
\end{align}
Since $[A_j,\texttt{Z}_i]=0$ for $j\neq i$ and $\{A_i,\texttt{Z}_i\}=0$, we have
\begin{equation}
    \{\texttt{Z}_i,\texttt{Z}_iA_iA_j\}=0,
\end{equation}
which implies
\begin{equation}
    \{\texttt{Z}_i,\texttt{Z}_iA_iC_i\}
    =0.
\end{equation}
Therefore,
\begin{align}
    \{\texttt{Z}_i,B\texttt{Z}_iB\}
    &=(2-k)\{\texttt{Z}_i,\texttt{Z}_i\}-2\{\texttt{Z}_i,\texttt{Z}_iA_iC_i\}\notag\\
    &=2(2-k)\mathbb{I}^{\otimes k}.
\end{align}
Consequently,
\begin{align}
    \{\texttt{Z}_i,U_{\rm Q}\texttt{Z}_iU^\dagger_{\rm Q}\}
    &=\frac{1}{L}\big(
    \{\texttt{Z}_i,\texttt{Z}_i\}
    +\{\texttt{Z}_i,[B,\texttt{Z}_i]\}
    -\{\texttt{Z}_i,B\texttt{Z}_iB\}
    \big)\notag\\
    &=\frac{2(k-1)}{L}\mathbb{I}^{\otimes k}.
\end{align}
Then,
\begin{equation}
    (\texttt{Z}_i+U_{\rm Q}\texttt{Z}_iU^\dagger_{\rm Q})^2
    =2\mathbb{I}^{\otimes k}+\{\texttt{Z}_i,U_{\rm Q}\texttt{Z}_iU^\dagger_{\rm Q}\}
    =\frac{4k}{L}.
\end{equation}
Hence,
\begin{equation}
    \xi_i^2=\mathbb{I}^{\otimes k}.
\end{equation}
Next consider the case $i=L-1$.
Similarly, we use
\begin{equation}
    B\texttt{Z}^{\otimes k}=-\texttt{Z}^{\otimes k}B.
\end{equation}
Then,
\begin{equation}
    B\texttt{Z}^{\otimes k}B
    =-B^2\texttt{Z}^{\otimes k}
    =k\texttt{Z}^{\otimes k}.
\end{equation}
Therefore,
\begin{align}
    U_{\rm Q}\texttt{Z}^{\otimes k}U^\dagger_{\rm Q}
    &=\frac{1}{L}\big((1-k)\texttt{Z}^{\otimes k}-2\texttt{Z}^{\otimes k}B\big).
\end{align}
Hence,
\begin{align}
    \{\texttt{Z}^{\otimes k},U_{\rm Q}\texttt{Z}^{\otimes k}U^\dagger_{\rm Q}\}
    &=\frac{1}{L}\big((1-k)\{\texttt{Z}^{\otimes k},\texttt{Z}^{\otimes k}\}-{2\{\texttt{Z}^{\otimes k},\texttt{Z}^{\otimes k}B\}}\big)\notag\\
    &=\frac{2(1-k)}{L}\mathbb{I}^{\otimes k}.
\end{align}
Thus,
\begin{equation}
    (\texttt{Z}^{\otimes k}-U_{\rm Q}\texttt{Z}^{\otimes k}U^\dagger_{\rm Q})^2
    =2\mathbb{I}^{\otimes k}-\{\texttt{Z}^{\otimes k},U_{\rm Q}\texttt{Z}^{\otimes k}U^\dagger_{\rm Q}\}
    =\frac{4k}{L}.
\end{equation}
Therefore,
\begin{equation}
    \xi_i^2=\mathbb{I}^{\otimes k}.
\end{equation}
Hence, the eigenvalues of $\xi_i$ are $\pm1$ for all $i\in[L]$.
Therefore, $\{E_i^0,E_i^1\}$ is a valid POVM for all $i\in[L]$.

We now compute the decoding success probability for these states and POVMs.
First consider the case $i<L-1$.
Since
\begin{equation}
    \langle{\bf c}({\bf b})|\texttt{Z}_i|{\bf c}({\bf b})\rangle=(-1)^{b_i}
\end{equation}
we have
\begin{align}
    \langle{\bf c}({\bf b})|U^\dagger_{\rm Q}\texttt{Z}_iU_{\rm Q}|{\bf c}({\bf b})\rangle
    &=\frac{1}{L}\big((k-1)\langle{\bf c}({\bf b})|\texttt{Z}_i|{\bf c}({\bf b})\rangle\notag\\
    &\quad-2{\langle{\bf c}({\bf b})|\texttt{Z}_iA_i|{\bf c}({\bf b})\rangle\big)}\notag\\
    &=\frac{L-2}{L}(-1)^{b_i}.
\end{align}
Therefore,
\begin{equation}
    \langle\psi({\bf b})|\xi_i|\psi({\bf b})\rangle
    =\sqrt{\frac{L-1}{L}}(-1)^{b_i}.
\end{equation}
Hence,
\begin{align}
    \operatorname{tr}\big(E^{b_i}_{i}\rho_{\rm Q}({\bf b})\big)
    &=\frac{1}{2}+(-1)^{b_i}\frac{1}{2}\langle\psi({\bf b})|\xi_i|\psi({\bf b})\rangle\notag\\
    &=\frac{1}{2}+\frac{1}{2}\sqrt{\frac{L-1}{L}}.
\end{align}
Next consider the case $i=L-1$.
Since
\begin{equation}
    \langle{\bf c}({\bf b})|\texttt{Z}^{\otimes k}|{\bf c}({\bf b})\rangle=(-1)^{p({\bf c})},
\end{equation}
we obtain
\begin{align}
    \langle{\bf c}({\bf b})|U^\dagger_{\rm Q}\texttt{Z}^{\otimes k}U_{\rm Q}|{\bf c}({\bf b})\rangle
    &=\frac{1}{L}\big((1-k)\langle{\bf c}({\bf b})|\texttt{Z}^{\otimes k}|{\bf c}({\bf b})\rangle\notag\\
    &\quad-2{\langle{\bf c}({\bf b})|\texttt{Z}^{\otimes k}B|{\bf c}({\bf b})\rangle\big)}\notag\\
    &=\frac{2-L}{L}(-1)^{p({\bf c})}
\end{align}
Noting that
\begin{equation}
    \langle{\bf c}({\bf b})|U^\dagger_{\rm Q}\xi_i U_{\rm Q}|{\bf c}({\bf b})\rangle=-\langle{\bf c}({\bf b})|\xi_i|{\bf c}({\bf b})\rangle,
\end{equation}
we have
\begin{align}
    \langle\psi({\bf b})|\xi_i|\psi({\bf b})\rangle
    &=(-1)^{p({\bf b})}\langle{\bf c}({\bf b})|\xi_i|{\bf c}({\bf b})\rangle\notag\\
    &=(-1)^{p({\bf b})+p({\bf c})}\sqrt{\frac{L-1}{L}}\notag\\
    &=(-1)^{b_i}\sqrt{\frac{L-1}{L}}.
\end{align}
Therefore,
\begin{align}
    \operatorname{tr}\big(E^{b_i}_{i}\rho_{\rm Q}({\bf b})\big)
    &=\frac{1}{2}+(-1)^{b_i}\frac{1}{2}\langle\psi({\bf b})|\xi_i|\psi({\bf b})\rangle\notag\\
    &=\frac{1}{2}+\frac{1}{2}\sqrt{\frac{L-1}{L}}
\end{align}
for all ${\bf b}\in\{0,1\}^L$ and $i\in[L]$.

As a result, both the average and worst-case decoding success probabilities are
\begin{equation}
    \frac{1}{2}+\frac{1}{2}\sqrt{\frac{L-1}{L}},
\end{equation}
which attains the value of the conjectured bound (Eq.~\eqref{eq:qrac-conjecture}).
\end{proofE}

\begin{figure*}[tbh]
    \centering
    \includegraphics[scale=1]{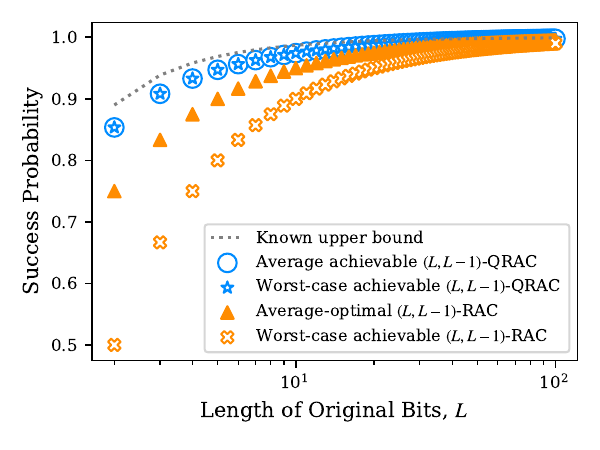}
      \caption{Analytically obtained decoding success probability of $(L,L-1)$-(Q)RACs, together with the known upper bound~\cite{ambainis1999dense,nayak1999optimal}.}
  \label{fig:ll1}
\end{figure*}

Figure~\ref{fig:ll1} plots the decoding success probabilities of the proposed $(L,L\!-\!1)$-(Q)RACs.
If the two conjectures (Eq.~\eqref{eq:qrac-conjecture} and Conjecture~\ref{conjecture:worst}) hold, then this plot represents the gap between the classical and quantum optimal values.

\subsection{Nonlinearity of Optimal RACs}
\label{sec:}

Finally, we briefly comment on the nonlinearity of optimal RACs.
We have shown that, for several families of $(L,k)$, optimal RACs are realized by well-known infinite families of linear codes.
However, an optimal RAC is not necessarily linear in general.
The examples are as follows:

\begin{remarkE}[][normal]
\label{remark:rac-code}
The average-optimal $(12,3)$-RAC is non-linear.
Indeed,
\begin{equation}
    \frac{1}{2^L}
    \min_{\substack{C\le\V\\ \dim(C)=k}}
    \sum_{{\bf b}\in\V}
    \min_{{\bf c}\in C}
    d_{\rm H/L}({\bf b},{\bf c})
    =
    \frac{13600}{12\cdot2^{12}},
\end{equation}
where the minimum is taken over all linear codes $C$. However, there exists a non-linear code
\begin{dmath}
    S=\{
    \texttt{000000000000}, \allowbreak
    \texttt{010111001110}, \allowbreak
    \texttt{100011111101}, \allowbreak
    \texttt{011100111101}, \allowbreak
    \texttt{101101011011}, \allowbreak
    \texttt{000110110010}, \allowbreak
    \texttt{111001100110}, \allowbreak
    \texttt{111010000001}\}
\end{dmath}
such that
\begin{equation}
    \frac{1}{2^L}
    \sum_{{\bf b}\in\V}
    \min_{{\bf s}\in S}
    d_{\rm H/L}({\bf b},{\bf s})
    =
    \frac{13599}{12\cdot2^{12}}.
\end{equation}
Similarly, the worst-case optimal $(11,4)$-RAC is also non-linear. For linear codes,
\begin{equation}
    \min_{\substack{C\le\V\\ \dim(C)=k}}
    \max_{{\bf b}\in\V}
    \min_{{\bf y}\in \operatorname{conv}(C)}
    d_\infty({\bf b},{\bf y})
    =
    0.4,
\end{equation}
whereas there exists a non-linear code
\begin{dmath}
    S=\{
    \texttt{00000000000}, \allowbreak
    \texttt{00011111111}, \allowbreak
    \texttt{01100001111}, \allowbreak
    \texttt{01111110000}, \allowbreak
    \texttt{00100110011}, \allowbreak
    \texttt{00111001100}, \allowbreak
    \texttt{01001010101}, \allowbreak
    \texttt{01010101010}, \allowbreak
    \texttt{11100111100}, \allowbreak
    \texttt{11111000011}, \allowbreak
    \texttt{10001101001}, \allowbreak
    \texttt{10010010110}, \allowbreak
    \texttt{11001011010}, \allowbreak
    \texttt{11010100101}, \allowbreak
    \texttt{10101100110}, \allowbreak
    \texttt{10110011001}\}
\end{dmath}
satisfying
\begin{equation}
    \max_{{\bf b}\in\V}
    \min_{{\bf x}\in \operatorname{conv}(S)}
    d_\infty({\bf b},{\bf x})
    =
    0.375.
\end{equation}
\end{remarkE}

\section{Numerical Experiments}
\label{sec:numerical}
\subsection{Average-Optimal / Worst-Case Achievable \texorpdfstring{$(L,k)$}{(L,k)}-RAC}

\begin{table*}[tbh]
\centering
\caption{Optimal average success probability of $(L,k)$-RAC obtained by solving the problem Eq.~\eqref{eq:average-prob}. The corresponding closed-form upper bounds (Eq.~\eqref{eq:average}) are also provided for comparison. ``$=$'' indicates the same value as the optimal success probability.}
\label{tab:rac-average}
\renewcommand{\arraystretch}{1.2}
\begin{tabular}{crr}
\toprule
$(L,k)$ & Optimal Success Prob. & Closed-form U.B. (Eq.~\eqref{eq:average})\\
\midrule
$(2,1)$ & $0.75$ & $=$ \\
$(3,1)$ & $0.75$ & $=$ \\
$(3,2)$ & $0.8\overline{3}=\frac{5}{6}$ & $=$ \\
$(4,1)$ & $0.6875$ & $=$ \\
$(4,2)$ & $0.8125$ & $=$ \\
$(4,3)$ & $0.875$ & $=$ \\
$(5,1)$ & $0.6875$ & $=$ \\
$(5,2)$ & $0.775$ & $=$ \\
$(5,3)$ & $0.85$ & $=$ \\
$(5,4)$ & $0.9$ & $=$ \\
$(6,1)$ & $0.65625=\frac{21}{32}$ & $=$ \\
$(6,2)$ & $0.75=\frac{3}{4}$ & $=$ \\
$(6,3)$ & $0.8\overline{3}=\frac{5}{6}$ & $=$ \\
$(6,4)$ & $0.875=\frac{7}{8}$ & $=$ \\
$(6,5)$ & $0.91\overline{6}=\frac{11}{12}$ & $=$ \\
$(7,1)$ & $0.65625=\frac{21}{32}$ & $=$ \\
$(7,2)$ & $0.7276\overline{785714}=\frac{163}{224}$ & ${\bf 0.7410\overline{714285}=\frac{83}{112}}$ \\
$(7,3)$ & $0.7946\overline{428571}=\frac{89}{112}$ & $=$ \\
$(7,4)$ & $0.875=\frac{7}{8}$ & $=$ \\
$(7,5)$ & $0.89\overline{285714}=\frac{25}{28}$ & $=$ \\
$(7,6)$ & $0.9\overline{285714}=\frac{13}{14}$ & $=$ \\
$(8,1)$ & $0.63671875=\frac{163}{256}$ & $=$ \\
$(8,2)$ & $0.716796875=\frac{367}{512}$ & $=$ \\
$(8,3)$ & $0.77734375=\frac{199}{256}$ & ${\bf 0.7890625=\frac{101}{128}}$ \\
$(8,4)$ & $0.828125=\frac{53}{64}$ & $=$ \\
$(8,5)$ & $0.890625=\frac{57}{64}$ & $=$ \\
$(8,6)$ & $0.90625=\frac{29}{32}$ & $=$ \\
$(8,7)$ & $0.9375=\frac{15}{16}$ & $=$ \\
\bottomrule
\end{tabular}
\end{table*}

\begin{table*}[tbh]
\centering
\caption{Achievable worst-case decoding success probability of RAC obtained by solving the problem Eq.~\eqref{eq:min-hausdorff-reduced}. For the case $(7,5)$, the problem in \eqref{eq:min-hausdorff-reduced} could not be fully solved due to a timeout. Nevertheless, we have confirmed that the optimal value lies within the interval $[0.75,\,0.79766]$.}
\label{tab:rac-worst}
\renewcommand{\arraystretch}{1.2}
\begin{tabular}{cr}
\toprule
$(L,k)$ & Achievable Prob.\\
\midrule
$(3,2)$ & $0.\overline{6}=\frac{2}{3}$\\
$(4,3)$ & $0.75$ \\
$(5,3)$ & $0.\overline{6}=\frac{2}{3}$ \\
$(5,4)$ & $0.8$ \\
$(6,3)$ & $0.\overline{6}=\frac{2}{3}$ \\
$(6,4)$ & $0.75$ \\
$(6,5)$ & $0.8\overline{3}=\frac{5}{6}$ \\
$(7,3)$ & $0.\overline{571428}=\frac{4}{7}$ \\
$(7,4)$ & $0.75$ \\
$(7,5)$ & $0.75^*$ \\
$(7,6)$ & $0.\overline{857142}=\frac{6}{7}$ \\
\bottomrule
\end{tabular}
\end{table*}

In this section, for small values of $L$ and $k$, we report the average-optimal $(L,k)$-RACs as well as worst-case achievable $(L,k)$-RACs.
We solved the MILPs Eq.~\eqref{eq:milp-average} and Eq.~\eqref{eq:milp-worst} using Gurobi~\cite{gurobi}.
The results are summarized in Tables~\ref{tab:rac-average} and~\ref{tab:rac-worst}.
Although a worst-case optimal RAC could, in principle, be found via a branch-and-bound search, the computation becomes prohibitively expensive even for $L=3$; therefore, we report only the achievable solutions obtained by the MILP.
Table~\ref{tab:rac-average} also lists the closed-form upper bound given in Eq.~\eqref{eq:average} for comparison.

The $k=1$ results in Table~\ref{tab:rac-average} coincide with the average-optimal $(L,1)$-RACs reported in~\cite{ambainis2024quantum}, providing a consistency check.

As can be seen from Table~\ref{tab:rac-average}, the closed-form upper bound in Eq.~\eqref{eq:average} is quite tight for $L\le 8$.
A nonzero gap from the optimum was observed only for $(L,k)=(7,2)$ and $(L,k)=(8,3)$.

For the average-optimal $(7,2)$-RAC, one minimizer of \eqref{eq:average-prob} is, for example,
\[
S=\{\texttt{0000100},\;\texttt{0111001},\;\texttt{1010111},\;\texttt{1101010}\}.
\]
Recall that the derivation of the bound in \eqref{eq:average} relies on the following implicit condition:
unless a string ${\bf b}$ belongs to the set of farthest points from $S$, the nearest element of $S$ is assumed to be unique.
In the present case, the farthest points from $S$ (e.g., \texttt{0000011}) have Hamming distance $3$.
However, there exist $12$ strings (including \texttt{0000111}) whose Hamming distance from $S$ equals $2$ and hence they do not belong to the farthest layer, yet they have two nearest bit strings in $S$ (e.g., \texttt{0000100} and \texttt{1010111}).
This multiplicity leads to double counting in the layer-size argument used in \eqref{eq:average}, which in turn underestimates the number of distance-$3$ points.
As a consequence, Eq.~\eqref{eq:average} yields a value larger than the true (tight) upper bound, i.e., it becomes a loose bound in such cases.

\subsection{Comparison of \texorpdfstring{$(L,k)$}{(L,k)}-RAC and \texorpdfstring{$(L,k)$}{(L,k)}-QRAC}
For QRACs, we implemented in PyTorch an optimization problem in which the density operators and POVMs were treated as trainable parameters, and the negative decoding success probability was used as the loss function.
The optimization was then carried out by gradient-based methods.
To mitigate convergence to poor local optima, we injected noise and re-optimized the parameters so as to escape shallow local minima.
When solving the problem of maximizing the worst-case decoding success probability, we replaced the nondifferentiable $\min$ operation with a softmin approximation.
For the case $k=L-1$, we also constructed density operators by the method of Theorem~\ref{theorem:optimal-qrac}.
In Table~\ref{tab:qrac-average}, these constructions are denoted by $(\bullet)^{*}$.
In addition, we evaluated QRACs obtained as simple tensor products of optimal smaller QRACs. For example, for $(L,k)=(7,3)$, we considered constructions such as
\begin{equation}
    \rho(\mathbf{b})
    =
    \rho_{11}(b_6)\otimes\rho_{31}(b_5b_4b_3)\otimes\rho_{31}(b_2b_1b_0)
\end{equation}
and
\begin{equation}
    \rho(\mathbf{b})
    =
    \rho_{21}(b_6b_5)\otimes\rho_{21}(b_4b_3)\otimes\rho_{31}(b_2b_1b_0),
\end{equation}
where $\rho_{11}$, $\rho_{21}$, and $\rho_{31}$ denote the density operators of the optimal $(1,1)$-, $(2,1)$-, and $(3,1)$-QRACs, respectively.
In Table~\ref{tab:qrac-average}, these constructions are denoted by $(\bullet)^{\otimes}$.
Among all families of density operators obtained by the methods above, we report the ones that achieved the highest average and worst-case decoding success probabilities.
Note, however, that the average and worst-case decoding success probabilities listed in the table are not necessarily achieved by the same family of quantum states.
For example, for $(L,k)=(5,2)$, the family of quantum states attaining the worst-case decoding success probability $0.81100$ is different from the one attaining the average decoding success probability $0.81463$.
Finally, for $(L,k)$ with $L>3$ and $k=1$, the nonexistence of schemes whose worst-case decoding success probability exceeds $0.5$ has already been proved~\cite{iwama2007unbounded}.
For this reason, we do not include values for this regime in the table.

\begin{figure*}[t]
  \centering
    \includegraphics[scale=1]{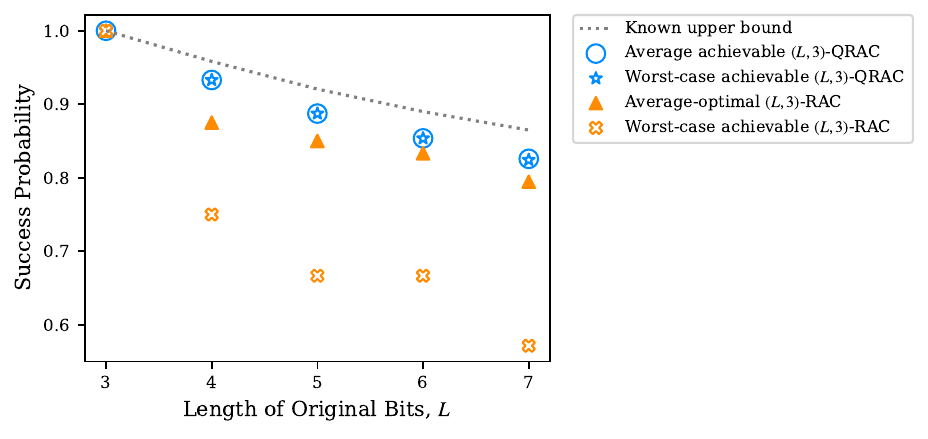}
    \label{fig:}%
  \caption{
  Numerically obtained decoding success probability of $(L,3)$-(Q)RACs, together with the known upper bound~\cite{ambainis1999dense,nayak1999optimal}.
  }
  \label{fig:L123}
\end{figure*}

Figures~\ref{fig:L123} plot the decoding success probabilities of $(L,k)$-(Q)RACs for $L\le7$ and $k=3$.
From this plot, we find that for at least $L\le 7$ there exist $(L,k)$-QRACs whose achieved average success probabilities are close to the conjectured bound.
Moreover, when judged by the average decoding success probability, the gap between RACs and QRACs is relatively small, consistent with prior observations~\cite{ambainis1999dense}.
In contrast, when focusing on the worst-case decoding success probability, the achievable RACs are substantially worse than the achievable QRACs.
In our plots, the worst-case success probability of the obtained QRACs is close to their average success probability, whereas for RACs (except for the trivial case $k=L$) the worst-case success probability is markedly smaller than the average success probability.

One possible explanation for this gap is that, due to computational constraints, in solving Eq.~\eqref{eq:worst-prob} we restricted the design set to $S\subset\{0,1\}^L$ rather than allowing $S\subset[0,1]^L$.
To test this possibility, we considered the relatively inexpensive case $(L,k)=(3,2)$ and computed a worst-case optimal RAC via a branch-and-bound search.
The resulting optimal set still satisfied $S\subset\{0,1\}^L$, and allowing $S\subset[0,1]^L$ did not improve the worst-case decoding success probability of the $(3,2)$-RAC.
Although our verification was limited to extremely small instances, these results suggest that the large gap between the average and worst-case decoding success probabilities for RACs is an intrinsic feature.
Equivalently, the most pronounced classical--quantum separation  in the non-asymptotic regime may occur in the worst-case decoding success probability.

\newif\ifusecolC
\usecolCtrue
\begin{table*}[tbh]
\centering
\caption{Achievable decoding success probabilities of QRACs. Here, $(\bullet)^*$ denotes values computed from the quantum states obtained via the construction in Theorem~\ref{theorem:optimal-qrac}, whereas $(\bullet)^{\otimes}$ denotes values computed from quantum states constructed by taking tensor products of QRACs with smaller $L$ and/or $k$.}
\label{tab:qrac-average}
\begin{tabular}{cll}
\toprule
$(L,k)$ & Worst-case & Average \\
\midrule
$(2,1)$ & $0.85355^*$ & $0.85355^*$ \\
$(3,1)$ & $0.78868$ & $0.78868$ \\
$(3,2)$ & $0.90825^*$ & $0.90825^*$ \\
$(4,1)$ & -- & $0.74148$ \\
$(4,2)$ & $0.85355^\otimes$ & $0.85355^\otimes$ \\
$(4,3)$ & $0.93301^*$ & $0.93301^*$ \\
$(5,1)$ & -- & $0.71358$ \\
$(5,2)$ & $0.81169$ & ${0.81463^\otimes}$ \\
$(5,3)$ & $0.88730$ & $0.88730$ \\
$(5,4)$ & $0.94721^*$ & $0.94721^*$ \\
$(6,1)$ & -- & $0.69405$ \\
$(6,2)$ & $0.78868^\otimes$ & $0.78868^\otimes$ \\
$(6,3)$ & $0.85355^\otimes$ & $0.85355^\otimes$ \\
$(6,4)$ & $0.90825^\otimes$ & $0.90825^\otimes$ \\
$(6,5)$ & $0.95644^*$ & $0.95644^*$ \\
$(7,1)$ & -- & $0.67864$ \\
$(7,2)$ & $0.73719$ & $0.76184$ \\
$(7,3)$ & $0.82463$ & ${0.82575^\otimes}$ \\
$(7,4)$ & ${0.85355^\otimes}$ & $0.87766$ \\
$(7,5)$ & $0.92257$ & $0.92258$ \\
$(7,6)$ & $0.96291^*$ & $0.96291^*$ \\
\bottomrule
\end{tabular}
\end{table*}

\section{Conclusion}
\label{sec:conclusion}
In this paper, we formulated the construction of classical RACs that are optimal with respect to average and worst-case decoding success probabilities as geometric optimization problems.
Under the average criterion, the problem reduces to selecting representatives in ${0,1}^L$, whereas under the worst-case criterion it becomes a minimax problem over point sets in $[0,1]^L$ with respect to a distance-like objective.
This formulation yields optimality results for several parameter families.
Although optimal RACs need not be linear in general, the analytically established optima in these families are attained by standard infinite families of binary linear codes.

The same geometric viewpoint also gives explicit classical--quantum separations. For every $L>1$, we constructed an explicit $(L,1)$-QRAC whose average decoding success probability strictly exceeds the optimal classical value.
For the family $(2^k-1,k)$, we gave a worst-case optimal classical RAC and an explicit QRAC with strictly larger worst-case decoding success probability.
For the family $(L,L-1)$, we constructed an average-case optimal classical RAC and showed that it is also worst-case optimal conditional on the stated conjecture.
We further obtained, by a closely analogous construction, an $(L,L-1)$-QRAC attaining the conjectured upper bound on the decoding success probability.

Finally, for small parameter regimes, we solved the resulting optimization problems numerically to obtain average-optimal and worst-case achievable RACs beyond the analytically proven infinite families, and compared them with numerically optimized QRACs.
These computations suggest the existence of further classical--quantum gaps.

\section*{Acknowledgment}
\label{sec:Acknowledgment}
This work is supported by MEXT Quantum Leap Flagship Program Grant Number JPMXS0118067285 and JPMXS0120319794. 

The authors acknowledge the assistance of OpenAI's ChatGPT (GPT-5.5) in the preparation of the proofs.
All mathematical arguments and proofs have been independently verified by the authors.
The authors take full responsibility for the correctness of the results and the final presentation of the proofs.

\bibliographystyle{unsrtnat} 
\bibliography{reference}

\end{document}